\documentclass{ws-mpla}

\def\noCutCCQEeff{100}
\def\vetoCCQEeff{54.8}
\def\timeCCQEeff{54.3}
\def\radiCCQEeff{45.0}
\def\enerCCQEeff{39.7}
\def\lemuCCQEeff{36.0}
\def\tsubCCQEeff{29.1}
\def\distCCQEeff{26.6}

\def\noCutCCQEeffNub{100}
\def\vetoCCQEeffNub{50.8}
\def\timeCCQEeffNub{50.4}
\def\radiCCQEeffNub{42.2}
\def\enerCCQEeffNub{37.0}
\def\lemuCCQEeffNub{35.1}
\def\tsubCCQEeffNub{31.3}
\def\distCCQEeffNub{29.6}

\def\noCutCCQEpur{39.0}
\def\vetoCCQEpur{36.8}
\def\timeCCQEpur{36.8}
\def\radiCCQEpur{37.4}
\def\enerCCQEpur{46.3}
\def\lemuCCQEpur{62.3}
\def\tsubCCQEpur{71.0}
\def\distCCQEpur{77.0}

\def\noCutCCQEpurNub{34.4}
\def\vetoCCQEpurNub{28.5}
\def\timeCCQEpurNub{28.6}
\def\radiCCQEpurNub{28.9}
\def\enerCCQEpurNub{39.0}
\def\lemuCCQEpurNub{52.2}
\def\tsubCCQEpurNub{59.3}
\def\distCCQEpurNub{61.8}

\newcommand{\bea}{
  \begin{eqnarray}}
  \newcommand{\eea}{
  \end{eqnarray}}
\newcommand{\uz}{\,\textrm{cos}\, \theta_\mu}
\usepackage{multirow}

\def\ep{\epsilon}

\def\th{\theta}

\def\si{\sigma}

\def\De{\Delta}

\def\Ph{\Phi}

\def\nubar{\bar{\nu}}
\def\tmu{T_\mu}
\def\nub{\bar{\nu}}

\def\to{\rightarrow}

\def\no{\nonumber}
\def\no{\nonumber}
\def\beq{\begin{eqnarray}}
  \def\eeq{\end{eqnarray}}

\def\uMeV{\mbox{MeV}}

\def\ueVt{\mbox{eV}^2}

\def\nue{\nu_e}
\def\nuebar{\bar{\nu}_e}
\def\numu{\nu_\mu}
\def\numubar{\bar{\nu}_\mu}
\def\numub{\bar{\nu}_\mu}

\def\mup{\mu^{+}}
\def\mum{\mu^{-}}
\def\piz{\pi^{0}}

\def\pip{\pi^{+}}
\def\pim{\pi^{-}}

\begin{document}

\markboth{J. Grange and T. Katori}
{Charged Current Quasi-Elastic Cross Section Measurement in MiniBooNE}

%%%%%%%%%%%%%%%%%%%%% Publisher's Area please ignore %%%%%%%%%%%%%%
\catchline{}{}{}{}{}
%%%%%%%%%%%%%%%%%%%%%%%%%%%%%%%%%%%%%%%%%%%%%%%%%%%%%%%%%%%%%%%%%%% 

\title{CHARGED CURRENT QUASI-ELASTIC CROSS SECTION MEASUREMENTS IN MINIBOONE}

\author{JOSEPH GRANGE
  \footnote{Present address: Argonne National Laboratory, Argonne, IL, 60439, U.S.A, grange@anl.gov}
}
\address{Department of Physics, University of Florida,\\
  Gainesville, FL 32611, U.S.A.\\
  jgrange@phys.ufl.edu
}

\author{TEPPEI KATORI
  \footnote{Present address: Queen Mary University of London, London, E1 4NS, U.K, t.katori@qmul.ac.uk}
}
\address{Laboratory for Nuclear Science, Massachusetts Institute of Technology,\\
  Cambridge, MA 02139, U.S.A.\\
  katori@mit.edu
}
\maketitle

\pub{Received (Day Month Year)}{Revised (Day Month Year)}

\begin{abstract}

  The neutrino-induced charged-current quasi-elastic (CCQE, $\nu_l+n\to l^-+p$ or $\bar\nu_l+p\to l^++n$) 
  interaction is the most abundant interaction around 1~GeV, 
  and it is the most fundamental channel to study neutrino oscillations. 
  Recently, MiniBooNE published both muon neutrino\cite{MB_CCQE}, 
  and muon anti-neutrino\cite{MB_ANTICCQE} double differential cross sections on carbon. 
  In this review, we describe the details of these analyses and include some historical remarks. 

\end{abstract}

\ccode{PACS: 11.30.Cp, 14.60.Pq, 14.60.St}

\section{MiniBooNE experiment}	

The Mini Booster Neutrino Experiment (MiniBooNE) (2002-2012) is designed to detect 
$\nue$ ($\nuebar$) appearance signals from $\numu$ ($\numubar$) beam 
in $\De m^2\sim 1\ueVt$ region through charged current quasi-elastic (CCQE) interactions. 

\beq
\numu&\stackrel{oscillation}{\longrightarrow}&\nue+n\to e^-+p~,\no\\
\numubar&\stackrel{oscillation}{\longrightarrow}&\nuebar+p\to e^++n~.\no
\eeq 

Confirmation of such an oscillation signal would indicate
 new physics beyond the Standard Model, such as sterile neutrinos.  
To test for the existence of exotic mixing with $\De m^2\sim 1\ueVt$, 
the MiniBooNE experiment was designed to observe neutrino interactions with E$_{\nu}\sim$800~MeV 
in a mineral oil based Cherenkov detector at a baseline L$\sim$500~m.  
The energy and baseline values are chosen so that the ratio $L/E_{\nu}$ matches the signal 
reported previously by the LSND experiment\cite{lsnd}.  This allows sensitivity to the same $\De m^2$ region under the two-neutrino massive oscillation model. 

% The dominant uncertainties for the oscillation analysis involve 
% the level of knowledge for the various neutrino interactions around 800~MeV, 
% which are almost entirely unconstrained by external data and particularly 
% so for interactions with a nuclear target such as carbon.  
% The most important interactions for understanding the 
% oscillation signal include the muon neutrino neutral-current (NC) 
% $\piz$ events background to the electron neutrino CCQE candidates and 
% the rate of muon neutrino CCQE events required for the overall normalisation of the signal.

%For the success of the experiment, it is important to understand CCQE interactions. 
In order to reliably predict the appearance of $\nue$ and $\nuebar$,
it is crucial to first understand in detail the rate and kinematics of $\numu$
and $\numubar$ interactions.  This is particularly true in a single-detector
experiment such as MiniBooNE, where these physics samples serve as an important
constraint.
%In the absence of a near detector, these physics
%samples serve as an important constrain the 
%This study has been done through muon-(anti)neutrino CCQE interactions. 
Assuming lepton universality, 
$\numu$ ($\numubar$) CCQE interactions should be identical to $\nue$ ($\nuebar$) CCQE interactions, 
with the exception of effects arising from the charged lepton mass. 
This supposedly simple interaction is the subject of this review article. 
Due to nuclear effects, we find rich nature in this interaction  
not previously considered in neutrino experiments.

% With a detailed flux prediction obtained using data from the HARP experiment~\cite{HARP}, 
% MiniBooNE has turned these important rate constraints on the various interactions background to the oscillation sample 
% into absolute cross sections for many kinematic configurations.  
% Many of these configurations are the first of their kind, 
% perhaps most notably the double-differential
% cross section in muon kinematics in charged-current quasi-elastic interaction.
%% which offers unprecedented insight into the behaviour of the muon in CCQE interactions.  

In this section we give an overview of the MiniBooNE experiment. 
In the following section (Sec.~\ref{sec:beam}), 
we discuss the neutrino beam, 
the heart of all neutrino cross section measurements. 
The general method of cross section analysis is described in Sec.~\ref{sec:xsec}, 
and Sec~\ref{sec:numuccqe} begins the discussion of the MiniBooNE
$\numu$ CCQE measurement. 
CCQE interactions with the $\nubar$-mode beam is discussed 
in the following section (Sec.~\ref{sec:numubarccqe}), and 
 Sec.~\ref{sec:combined} describes the combined result.  The 
conclusions follow.

\begin{figure}[tbp]
  \centerline{\psfig{file=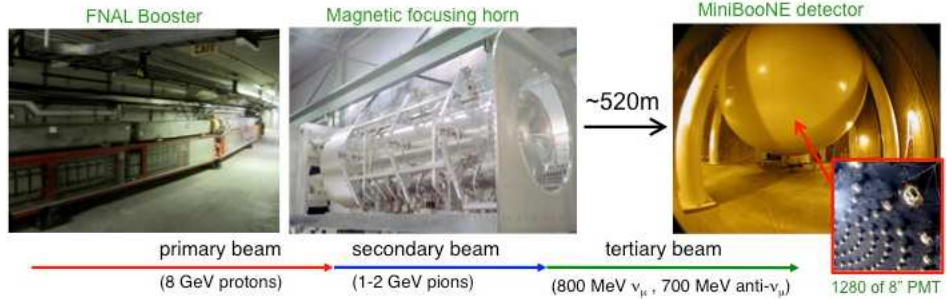,width=5.0in}}
  \vspace*{8pt}
  \caption{\label{fig:BNB} The scheme of the MiniBooNE experiment. 
    Protons (primary beam) are extracted from Fermilab Booster, 
    then send to the target located in the magnetic focusing horn to create shower of mesons (secondary beam). 
    Neutrinos are produced by decay-in-flight of mesons (tertiary beam), and detected by the MiniBooNE detector.}
\end{figure}

\subsection{Booster neutrino beamline (BNB)}

MiniBooNE accepts muon (anti)neutrino beam 
from the Fermilab Booster neutrino beamline (BNB)~\cite{MB_beam}. 
Fig.~\ref{fig:BNB} shows the scheme of the BNB.
The 8 GeV primary proton beam is extracted from the Booster 
and steered to collide with the beryllium target in the magnetic focusing horn. 
The collision of protons with the target creates a shower of secondary mesons, 
and the polarity of the surrounding toroidal field is chosen to focus $\pi^+$($\pi^-$) 
for $\nu$($\nubar$) mode.  
The horn simultaneously defocuses $\pi^-$($\pi^+$) to reduce the backgrounds 
from $\numubar$($\numu$) interactions in $\nu$($\nubar$) mode beam. 
The decay-in-flight of the sign-selected  $\pi^+$($\pi^-$) in an air-filled hall leads 
to a tertiary beam composed mostly of $\numu$($\numubar$).    
The decay length of typical pions are $\sim$18~m.
This wideband $\numu$($\numubar$) beam is peaked around 800 (650) MeV. 

\subsection{MiniBooNE detector}

The MiniBooNE detector is located 541~m north of the target\cite{MB_detec}. 
The detector is a 12.2~m diameter spherical Cherenkov detector filled 
with 800 tons of undoped mineral oil, whose chemical composition is dominantly CH$_{2}$. 
An inner region with diameter 11.5~m is covered with 1,280 8-inch PMTs (Fig.~\ref{fig:BNB}) and 
is optically separated from the 35~cm thick outer shell which houses 240 8-inch PMTs and acts as a veto 
to identify both exiting and entering charged particles. 

\begin{figure}[tbp]
  \centerline{
    \psfig{file=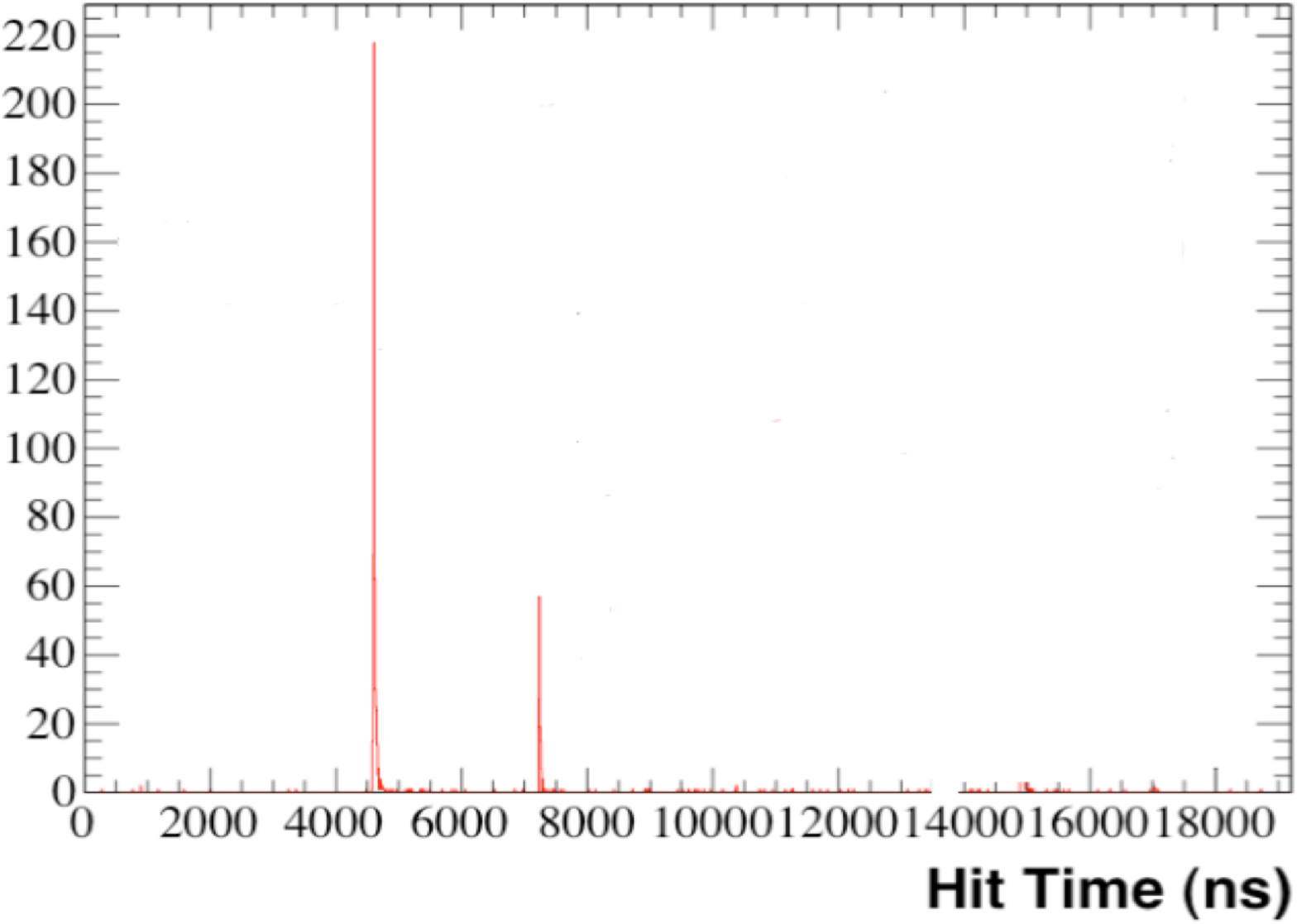,width=2.0in}
    \psfig{file=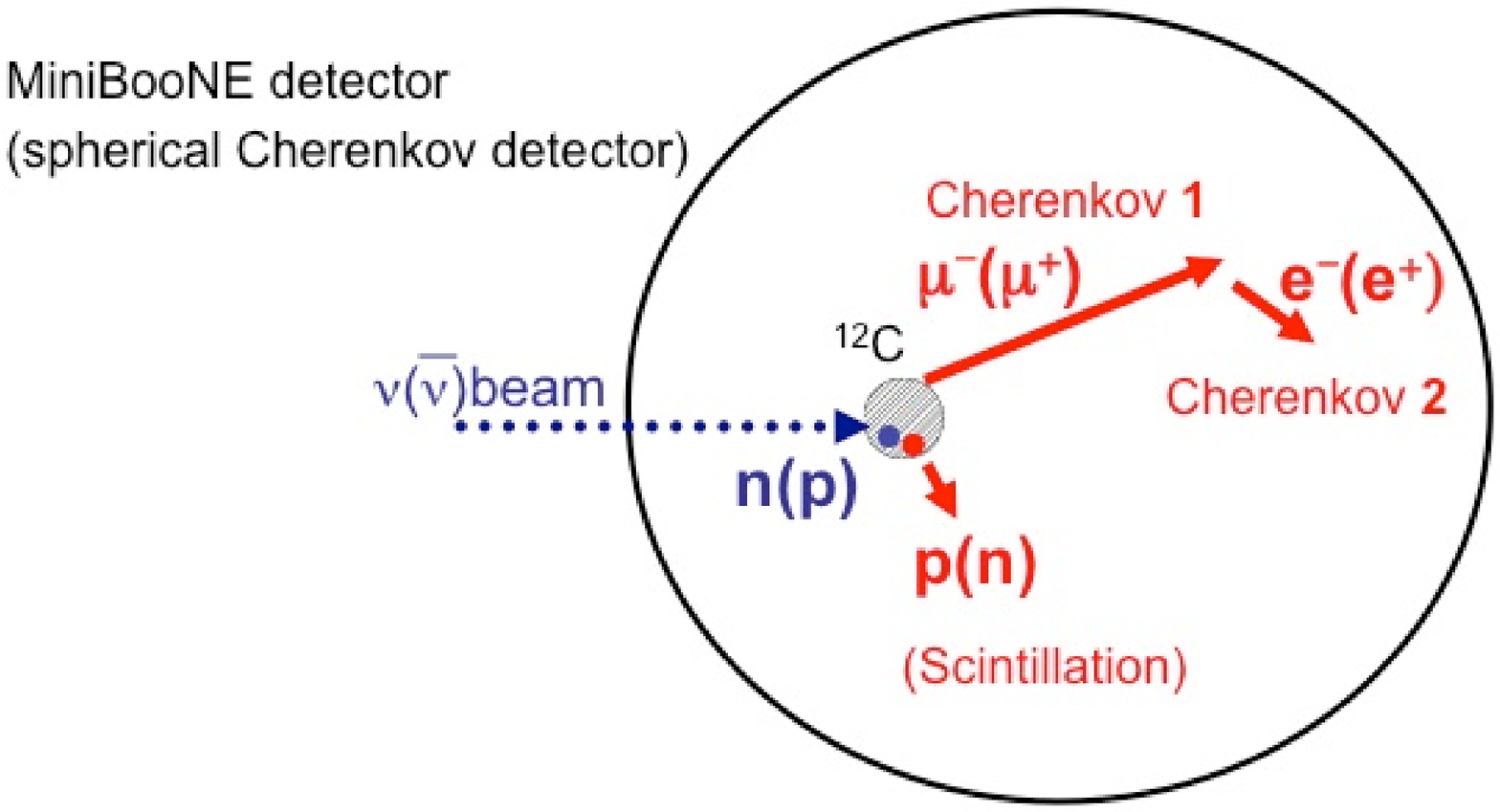,width=3.0in}
  }
  \vspace*{8pt}
  \caption{\label{fig:CCQEreaction}
    Left plot shows a data hit distribution for two subevents. The first high peak corresponds to muon Cherenkov ring, 
    and the subsequent small peak corresponds to the Cherenkov radiation from the electron following muon decay.   
    Right cartoon shows the CCQE interaction in MiniBooNE detector. 
    The muon (anti-muon) produces the primary Cherenkov ring, which is characterised with higher hits, 
    then subsequent decay produces weaker secondary Cherenkov ring from the electron (positron). 
    Nucleons are often below Cherenkov threshold and ignored. 
    Notice hydrogen atoms also participate in CCQE for muon anti-neutrino interactions.
  }
\end{figure}

The PMT timing information is used to associate clusters of activity with 
the signature of a single particle using PMT ``hits", and temporal groups of hits form ``subevents".  
%% A PMT pulse passing the discriminator threshold of $\sim$ 0.1 photoelectrons is defined as a hit, 
%% and is the basic unit of the observed signal intensity.  
%% A group of PMT activity with at least 10 hits within a 200~ns window and 
%% individual hit times less than 10~ns apart, 
%% while allowing for at most two spacings of 10 - 20~ns, defines a subevent.  
With high efficiency, subevents are used to identify and separate particles 
whose transit emits significant amounts of Cherenkov light and so are excellent 
for separating the signature and topology of muons from their decay electrons in CCQE interactions 
(Fig.~\ref{fig:CCQEreaction}, left).  

The primary result of the analyses described here is the flux-integrated differential cross section 
of muon (anti-muon) kinematics from the CCQE interaction. 

\beq
\numu+n\to\mu^-+p~,\no\\
\numubar+p\to\mu^++n~.\no
\eeq 

Figure~\ref{fig:CCQEreaction}, right, shows a cartoon of this reaction. 
We identify muon (anti-muon) by detecting first subevent from the muon (anti-muon) 
and delayed second subevent from the electron (positron). 
This two-fold signal defines the CCQE interaction. 
Details of signal definition and event selections are given in 
Sec.~\ref{subsec:definition} and Sec.~\ref{subsec:selection}.

A natural advantage of the MiniBooNE Cherenkov detector technology 
is its angular acceptance of the muon produced in CCQE interactions.  
The spherically symmetric geometry allows for equal angular acceptance over the full 4$\pi$ of solid angle. 
This is in contrast to forward type tracking detectors.  
The acceptance of tracking detectors is necessarily a function of the production angle, 
where muons created perpendicular to the neutrino direction are almost entirely missed, 
and backwards-going muons can be very challenging to reconstruct\cite{T2K_CCincl}.  
This is much more than just an experimental detail; 
the physics reach of a given detector is highly dependent on the angular acceptance.  
%% Broadly, and as described in more detail in Sec.~\ref{numuccqe}, 
%% forward-going muons are strongly correlated with interactions of low-momentum transfer 
%% which are known to be described poorly in most modern event generators, 
%% twhile two-body nuclear processes may contribute most to muons produced anti-parallel to the beam direction.  

\subsection{Event reconstruction}

The pattern, timing, and total charge of prompt Cherenkov radiation collected by the PMTs 
is used to identify particle type and its kinematics, which is the direct observable 
of the experiment and forms the basis of the differential cross sections~\cite{MB_recon}.  
For the $\numu$($\numubar$) CCQE cross section measurements described in this review, 
we identify muon tracks and their kinematics from CCQE interactions. 
To determine these crucial quantities, 
a likelihood function is compared to the topology and timing of the observed PMT hits:

\beq
\mathcal{L}({\bf x}) = \prod_{\textrm{unhit PMTs}\,\,i} \left(1 - P(i\,\textrm{hit};{\bf x}) \right) 
\times \prod_{\textrm{hit PMTs}\,\,i}P(i\,\textrm{hit};{\bf x})\,f_q(q_i;{\bf x})\,f_t(t_i;{\bf x}),
\label{eqn:lnLk}
\eeq

\noindent where $P(i\,\textrm{hit};{\bf x})$ is the probability for PMT $i$ to register 
a hit given the muon vertex and kinematic vector ${\bf x}$, and $f_q$ ($f_t$) 
is a probability distribution function (PDF) for the hit to return 
the measured charge (time) $q_i$ ($t_i$) operating under a muon hypothesis.  

The vector {\bf x} is composed of the muon time, 
energy and position at creation, 
as well as its momentum projections along the spherical azimuthal and polar angles. 
The negative logarithm of the likelihood function in Eq.~\ref{eqn:lnLk} simultaneously varies 
these seven parameters while comparing to the observed PMT hits.  
The parameters from the maximised likelihood function yield the reconstructed muon kinematics. 

Such likelihood function is also developed for other particles, and used for cross section measurements 
beyond the CCQE interactions, including neutral current $\piz$ production~\cite{MB_NCpi0}, 
neutral current elastic scattering~\cite{MB_NCEL,MB_ANTINCEL}, 
charged-current $\pip$ production~\cite{MB_CCpip}, 
and charged-current $\piz$ production~\cite{MB_CCpi0}. 
Indeed, MiniBooNE measured over 90\% of all possible 
 muon neutrino interactions in the MiniBooNE detector.

\section{Neutrino flux\label{sec:beam}}

Neutrino cross sections have been measured since 
the advent of the high intensity neutrino beam\cite{Sam_review}. 
The observable is the rate of the interaction. 
This can be described by the convolution of 
the neutrino flux ($\Ph$), 
the neutrino interaction cross section ($\si$), 
and the experimental detection efficiency ($\ep$). 

\beq
rate\propto\int\Ph\times\si\times\ep
\label{eq:xs}
\eeq

Clearly, we need to know the neutrino flux {\it a priori}
to infer the cross section from the measurement of neutrino rates in the detector. 
This is in general not easy, 
since high intensity modern neutrino beams are made from decay-in-flight 
mesons. 

Modern neutrino experiments typically rely heavily on 
the prediction of neutrino flux from simulations with varying degrees
of data constraints. 
Typical simulations include 
(1) primary proton beam propagation, and interaction with target material, 
(2) production of secondary mesons, propagation and decay, and 
(3) tertiary neutrino beam prediction under suitable geometry setting. 

\subsection{Primary proton interactions}

Historically, little attention is paid to the proton - target interaction model. 
However, there is a large uncertainty on this process, and even worse, 
those models are tuned from old and sparse data.
%Clearly, errors on this can and will be improved 
%in the future as cross-section measurements move
%into an era of greater precision.  
%subsequent to incorporating modern meson distributions into simulations, 
Such proton interaction model contributes the large uncertainty 
to both BNB~\cite{BNB_flux} and the T2K neutrino flux~\cite{T2K_flux}. 
Clearly, these error estimates can and will be improved 
in the future as cross-section measurements move
into an era of greater precision.  

\subsection{Meson production cross section}

\begin{figure}[tbp]
\centerline{
\psfig{file=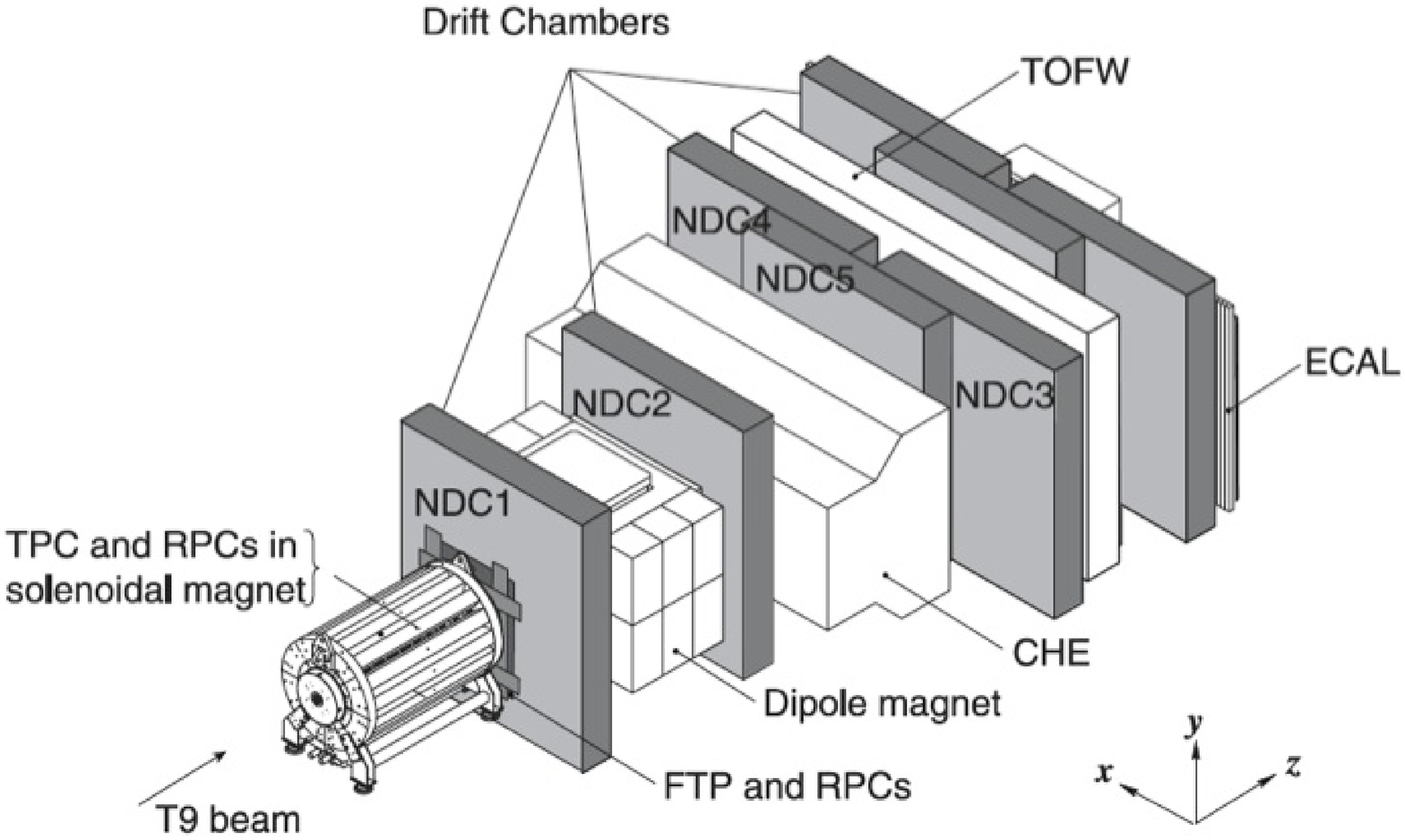,width=2.5in}
\psfig{file=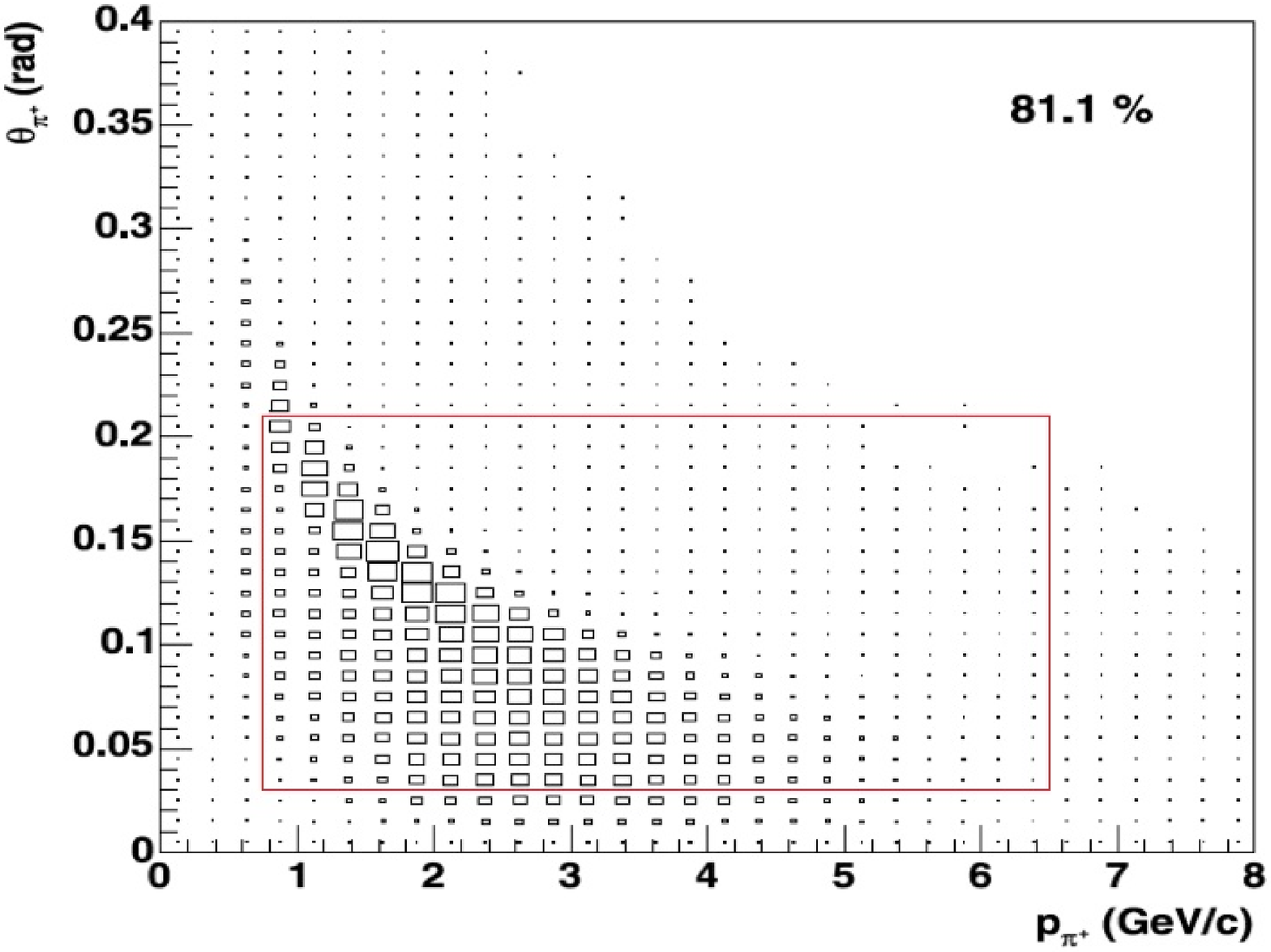,width=2.5in}
  }
\vspace*{8pt}
\caption{\label{fig:HARP} 
The left plot show the HARP detector. 
The detector consists of different sub-detectors to improve particle ID. 
The right plot shows the $p-\th$ distribution of $\pi^+$ in the BNB (simulation) 
which produces muon neutrinos passing through the MiniBooNE detector. 
}
\end{figure}

Since conventional high-intensity neutrino beams are made by decay-in-flight mesons, 
special attention is paid to the simulation of meson production. 
The program of choice for a reliable flux prediction involves 
dedicated external measurements of meson production, 
such as those provided by the HARP experiment for K2K and MiniBooNE.  
Figure~\ref{fig:HARP}, left, shows the drawing of the HARP detector~\cite{HARP}. 
%NA61(SHINE)~\cite{SHINE} for T2K, SPY~\cite{SPY} for NOMAD, and MIPP\cite{MIPP}. 
These hadroproduction experiments use tracking detectors to 
measure outgoing meson kinematics as precisely as possible.  
Figure~\ref{fig:HARP}, right, shows the $p-\th$ distribution 
of $\pi^+$ in the BNB (simulation) which produces muon neutrinos 
passing through the MiniBooNE detector. 
The red box shows the region measured by the HARP experiment. 
81.1\% of pions are directly measured, and outside of the box is extrapolated from the model. 

%%In the absence of such an option of dedicated hadroproduction data, 
%%the simulation of meson production is typically extrapolated from 
%%the available data to the relevant experimental proton energy and target material. 
%%Important to note, this prediction is clearly only as accurate 
%%as the input hadroproduction knowledge, 
%%and this is one reason why the uncertainty on a prediction
%%using data extrapolated from other energies and target material 
%%to the setup of interest is model-dependent and risky.

\subsection{Forward-going mesons}

Even with dedicated hadroproduction data, 
regions of meson phase space crucial to neutrino experiments may not be available.  
Since non-interacting protons are undeflected, hadroproduction detectors at
zero degrees relative to the proton beam become saturated and these regions 
are extremely challenging to measure. 
Therefore, mesons produced at very forward angles are often not reported, 
and neutrino experiments must either extrapolate the data into this region or find 
external constraints. 
For the HARP experiment, this corresponds to pions below 0.03~rad (Fig.~\ref{fig:HARP}, right).
%%Just as extrapolating non-dedicated hadroproduction data may be risky, 
%%such a kinematic extrapolation may not be accurate. 
As we will see in Sec.~\ref{subsec:forwardpi}, 
such a situation demanded MiniBooNE use a variety of {\it in situ} 
measurements to measure the $\pip$ production
cross section at low production angles to correct $\numu$ induced backgrounds in the $\nubar$-mode beam.

\subsection{Proton re-scattering}

The nuclear target size is chosen to maximise primary proton interaction rate, 
and an accompanying complication is the possibility of proton scattering more than once as it
traverses the material. The prediction of mesons produced in such interactions is known to be challenging, 
and unfortunately their contribution is typically not directly constrained by dedicated hadroproduction data.  
 
\begin{figure}[tbp]
\centerline{\psfig{file=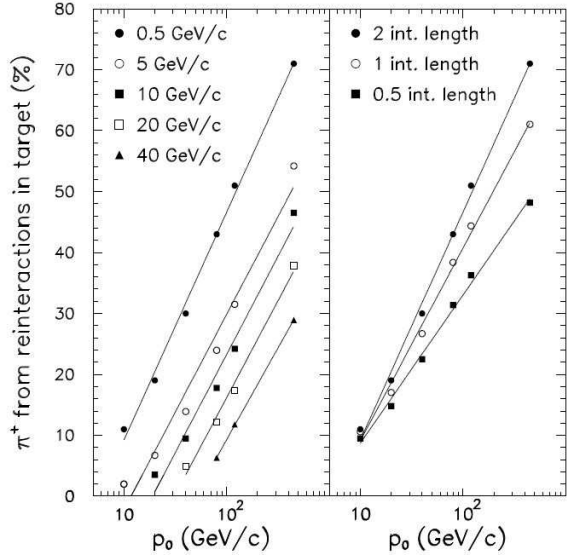,width=3.0in}}
\vspace*{8pt}
\caption{\label{fig:reInts} Simulation of the tertiary $\pip$ yield from re-interactions in a graphite target.  
Given as a function of incident proton beam momentum $p_0$, 
the $\pip$ fraction is given for the indicated thresholds 
on the longitudinal component of the $\pip$ momentum (left), 
and also for targets of 0.5, 1.0, and 2.0 interaction lengths (right). 
The primary proton beam at the BNB has momentum 8.9 GeV/c.}
\end{figure}

Hadroproduction experiments usually extract meson 
production cross sections using ``thin" nuclear targets,  
of proton interaction length $\sim$5\%, while the target used in 
the experiment is usually close to 2 interaction lengths. 
Thus a significant fraction of protons are scattered more than once to create mesons.   
%%{\bf verify this is also true for other expts?} 
%%In the case of MiniBooNE, The total cross section for these secondary interactions 
%%are calculated with the Glauber model~\cite{glauber}, 
%%and this calculation is verified with comparisons to data wherever possible.  
%%Based on the agreement between this model and the available data, 
%%uncertainties on the most important processes contributing pions to 
%%the beam are set around 20\% and higher~\cite{MB_beam}.  
%%Fortunately, while some details of this calculation are model-dependent, 
The left side of Figure~\ref{fig:reInts} shows\cite{zarko} 
the overall contribution of these processes 
to the total neutrino flux. 
For the BNB energy (8.9~GeV/c), this effect is rather low, 
and it is at the level of $\sim$ 10\%.  
The right side of the same figure also suggests the contribution from pions due to 
re-scattering protons present in the full MiniBooNE target 
but not in the thin target data collected by HARP is small.  
Therefore, with the exception of the very forward-going angular region addressed in Sec.~\ref{sec:numubarccqe}, 
the HARP data allows for a minimally model-dependent determination of 
the production of neutrino and anti-neutrino parent pions at the BNB. 
Note that proton re-scattering processes become more important with higher primary proton energy, 
such as MINERvA (using 120 GeV main injector), and T2K (30 GeV synchrotron ring). 

%%MiniBooNE used relatively lower energy beam (8 GeV Booster). 
%%This suppresses secondary mesons less than 10\% of total produced mesons. 

\subsection{Neutrino flux data driven correction}

As Eq.~\ref{eq:xs} shows, 
a measurement is a convolution of neutrino flux and neutrino cross section model. 
Therefore, it is dangerous to tune the neutrino flux from the neutrino data, 
and this was incorrectly done in past experiments. 
For example, it was common to model the neutrino flux from 
the CCQE interaction measurement in the same experiment\cite{Baker}, 
assuming the CCQE interaction model. 
Then, this tuned flux was used to measure the CCQE cross section..., 
of course you measure the cross section you assumed before! 
The danger of such procedure is, this obviously biases the measurement. 
The assumed of cross section model modifies the measured result. 
This is something we must avoid. For the data to be trusted for theorists to study their models, 
experimental data must be model independent as much as possible. 
As we will see in Sec.~\ref{subsec:mafit}, MiniBooNE CCQE data suggests 
no discrepancy is due to the originated from the flux model, 
and so we do not apply any data driven correction on the neutrino flux prediction 
(with the exception of forward going pions, as discussed in Sec.~\ref{subsec:forwardpi}).  
In a similar story, the roles of flux and cross section are swapped for the MINOS experiment. 
For MINOS, the majority of interactions are well-understood deep inelastic scattering (DIS). 
MINOS found the origin of data-simulation disagreement is from flux modeling, 
and meson productions are subsequently tuned in $P_T-P_z$ space by simultaneously fitting four different beam configurations\cite{zarko}.  The crucial issue is how reliable are the theoretical and experimental bases of the flux and cross section models.

\section{Neutrino cross section measurement\label{sec:xsec}}

%While the neutrino flux typically contributes the dominant uncertainty to measurements
%of absolute neutrino cross-sections, many other factors play important roles.  
As the community moves into an era of precision measurements through detector technology advancements
and the plentiful statistics afforded by high-intensity beams, an emphasis has been placed
on pushing the data collected into cross sections differential in as many distributions as possible.  This
allows for the most stringent test of the various predictions for processes possible for a given physics sample.  
For MiniBooNE, the main result of the CCQE analyses is the double-differential cross section in muon kinematics. 
In this section, we describe the cross section measurement method step-by-step. 
Eq.~\ref{eq:dsigma} shows an example of the differential 
cross section of muon kinetic energy from CCQE interaction. 

\beq
\frac{d\sigma}{d\tmu}_i = \frac{\sum_j U_{ij}\left(d_j - b_j\right)}{\Delta\tmu\ \,\epsilon_i \,\Phi \,T}, 
\label{eq:dsigma}
\eeq

Here, $d_{j}$ ($b_j$) is the data (background) reconstructed in the $j$th kinematic region of muon energy 
$T_\mu$, $U_{ij}$ is the probability 
for an event of true quantity within bin $i$ to be reconstructed in bin $j$, 
$\epsilon$ is the detection efficiency, $\Phi$ is the integrated neutrino flux and 
$T$ is the number of nuclear targets in the volume studied.  We discuss each in turn.

%%\begin{figure}[tbp]
%%\centerline{\psfig{file=xsec.eps,width=6.0in}}
%%\vspace*{8pt}
%%\caption{\label{fig:xsec} 
%%An overview of the differential cross section measurement of muon kinetic energy. 
%%}
%%\end{figure}

\subsection{Signal definition\label{subsec:definition}}
 
It is very important to define what is the ``signal'' of the measurement precisely. 
It is critical especially when we want to compare the results from other experiments or thoeries. 
So far, we call our signal to be``CCQE'' interaction, as Fig.~\ref{fig:CCQEreaction} right shows. 
Since MiniBooNE does not have a magnetic field, 
there is no charge separations and the signature from muons and anti-muons are degenerate on 
an event-by-event basis. 
We can discriminate charged pions by utilising decay products, 
and also neutral pions by detecting electromagnetic showers. 
However most protons are below Cherenkov threshold and we do not discriminate. 
Therefore, signal topology is defined to be ``1 muon + 0 pions and any number of protons''. 
 
However, there it is possible for non-CCQE channels to make this topology. 
Especially, when pions are absorbed in the target nuclei, 
CC pion production channels have intrisically same topology with CCQE. 
Therefore, pion production channels with nuclear pion absorption is referred ``irreducible background'', 
and the topology presented above can be called ``CCQE-like'' sample. 
After subtracting the irreducible background, data is called ``CCQE'' sample. 

These terminologies are MiniBooNE-specific and readers should pay attention to how 
CCQE is defined in other experiments. 
In fact, it was later mentioned two-body current 
interaction also contributes an irreducible background.  However, 
we did not subtract them from the CCQE-like sample because 
we such prediction was not available at the time the analyses were performed. 

\subsection{Physics sample selection, $d_j$\label{subsec:selection}}

The aim for the selection of any physics sample is to retain as many high-quality signal events 
while rejecting as much background as possible.  
Table~\ref{tab:sel} lists the requirements of the physics sample along with purity and 
detection efficiency figures for both $\numu$ and $\numub$ CCQE-like data sample.  
An important difference between the two lies in the purity, 
where $\numu$ interactions contribute significantly to the $\numub$ sample but not {\it vice versa}.  
It can also be seen that the $\numub$ efficiency is around 10\% higher relative to 
the $\numu$ case subsequent to the requirement of two and only two observed subevents.  
The $\mum$ from $\numu$ events not accepted by this cut have undergone nuclear capture, 
which is an unavoidable loss of $\numu$ CCQE signal events on nuclear material when requiring the presence of the electron from $\mum$ decay.  

While all CC $\numu$ detection efficiency suffers from this nuclear capture, 
the lack of such interactions between $\mup$ and nuclear material can be exploited 
as a tool to provide discrimination between $\numu$ and $\numub$ CC interactions in the absence of a magnetic field. 
This novel technique was first demonstrated by MiniBooNE, 
and is described along with other such analyses in Sec.~\ref{subsec:forwardpi}.  
Detailed understanding and exploitation of $\mum$ nuclear capture is also important 
for the future of precision neutrino oscillation experiments, 
where the prevalent use of liquid argon detectors expose $\mum$'s to a probability for nuclear capture\cite{arCap} of $\sim$70\%.

\begin{table}
\tbl{Selection efficiency ($\ep$) and purity (pur.) for the $\numu$ and $\numub$ CCQE samples.  
Requirements are based on well-understood lepton kinematics, 
and so sample selection does not suffer from the interaction model dependence.}
{\begin{tabular}{lp{0.6\textwidth}cccc}\toprule
\hline
\hline
cut&\multirow{2}{*}{description} &\multicolumn{2}{c}{$\numu$ CCQE}&\multicolumn{2}{c}{$\numub$ CCQE}   \\
\# & &$\epsilon$  & pur.&$\epsilon$  & pur.\\
\hline
0& no cuts &$\noCutCCQEeff$&$\noCutCCQEpur$&$\noCutCCQEeffNub$&$\noCutCCQEpurNub$ \\
1& all subevents, \# of veto hits $<6$                         		 &$\vetoCCQEeff$ & $\vetoCCQEpur$&$\vetoCCQEeffNub$ & $\vetoCCQEpurNub$ \\
2& 1st subevent, event time window, \newline $4400<T(\mathrm{ns})<6400$ &$\timeCCQEeff$ & $\timeCCQEpur$&$\timeCCQEeffNub$ & $\timeCCQEpurNub$ \\
3& 1st subevent, reconstructed vertex \newline radius $<500$~cm         &$\radiCCQEeff$ & $\radiCCQEpur$&$\radiCCQEeffNub$ & $\radiCCQEpurNub$ \\
4& 1st subevent, kinetic \newline energy $>200$~MeV                     &$\enerCCQEeff$ & $\enerCCQEpur$&$\enerCCQEeffNub$ & $\enerCCQEpurNub$ \\
5& 1st subevent, $\mu/e$ log-likelihood ratio $>0.0$           &$\lemuCCQEeff$ & $\lemuCCQEpur$&$\lemuCCQEeffNub$ & $\lemuCCQEpurNub$ \\
6& 2 subevents	                                                            &$\tsubCCQEeff$ & $\tsubCCQEpur$ &$\tsubCCQEeffNub$ & $\tsubCCQEpurNub$ \\
7& 1st subevent, $\mu-e$ vertex distance $>100$~cm and &                          \\
      & $\mu-e$ vertex distance $>(500 \times T_\mu(\mathrm{GeV})-100$)~cm &$\distCCQEeff$ & $\distCCQEpur$&$\distCCQEeffNub$ & $\distCCQEpurNub$ \\
\hline
\hline
\end{tabular}}
\label{tab:sel}
\end{table}

Broadly, this selection is entirely based on simple kinematic observations of the Cherenkov light 
from prompt muon and it's decay electron to ensure the CCQE events. 
%studied are well-reconstructed while also optimizing the detection efficiency and accepting background interactions as low as feasible. 
An immediate benefit to this rather simple selection is the absence of requirements on hadronic activity. 
While the physics produced from studies of this sample cannot test the kinematics of hadronic behavior in CCQE interactions, 
the selection avoids any model-dependence through assumptions of proton and neutron production and their coupling to instrumental thresholds.  
This is a complication common to all tracking detectors. 
Even with a perfect tracking detector, 
final-state interactions severely complicate the interpretation of such hadronic observations.

\subsection{Background subtraction, $d_j-b_j$}

\begin{figure}[tbp]
\centerline{
\psfig{file=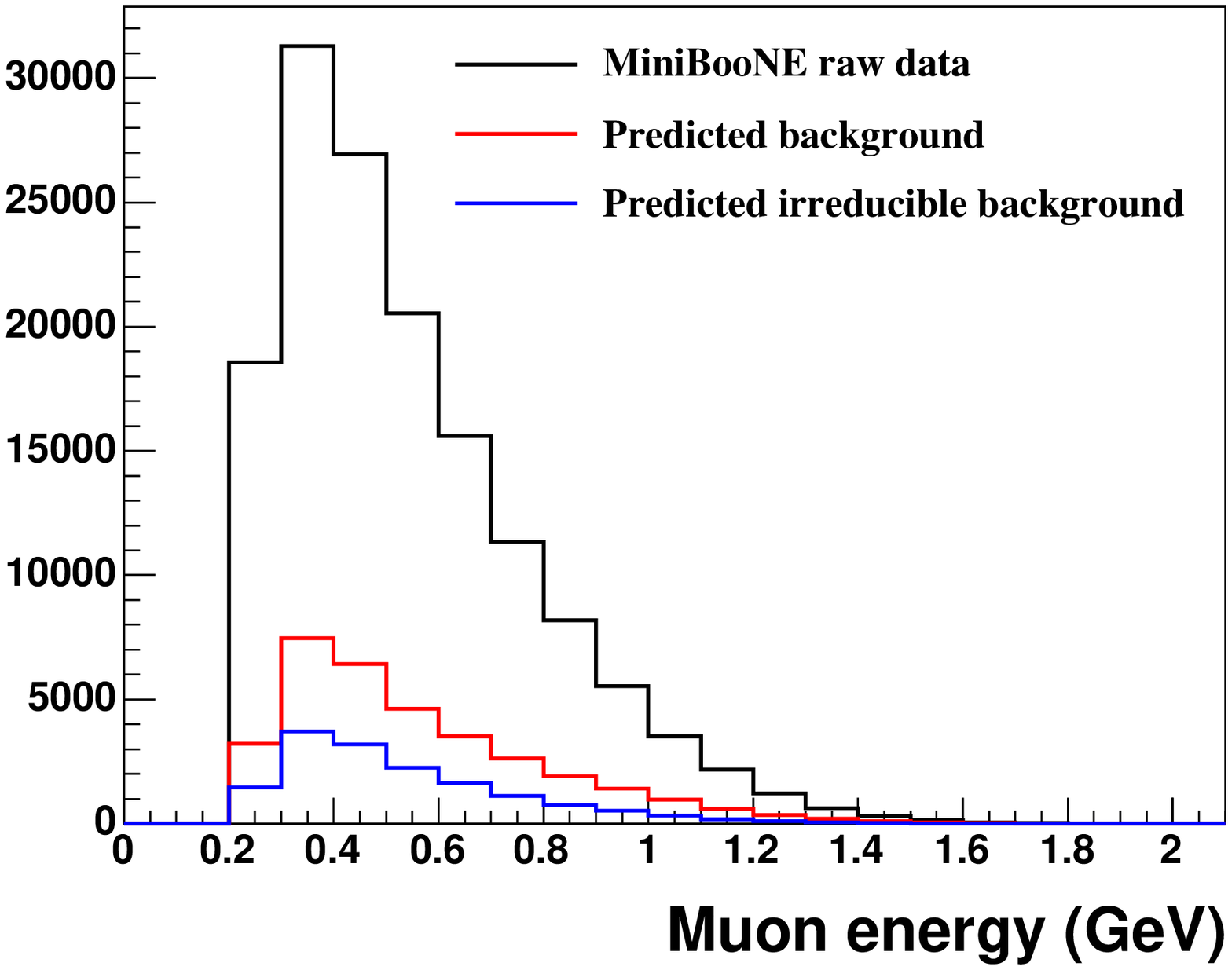,width=2.5in}
\psfig{file=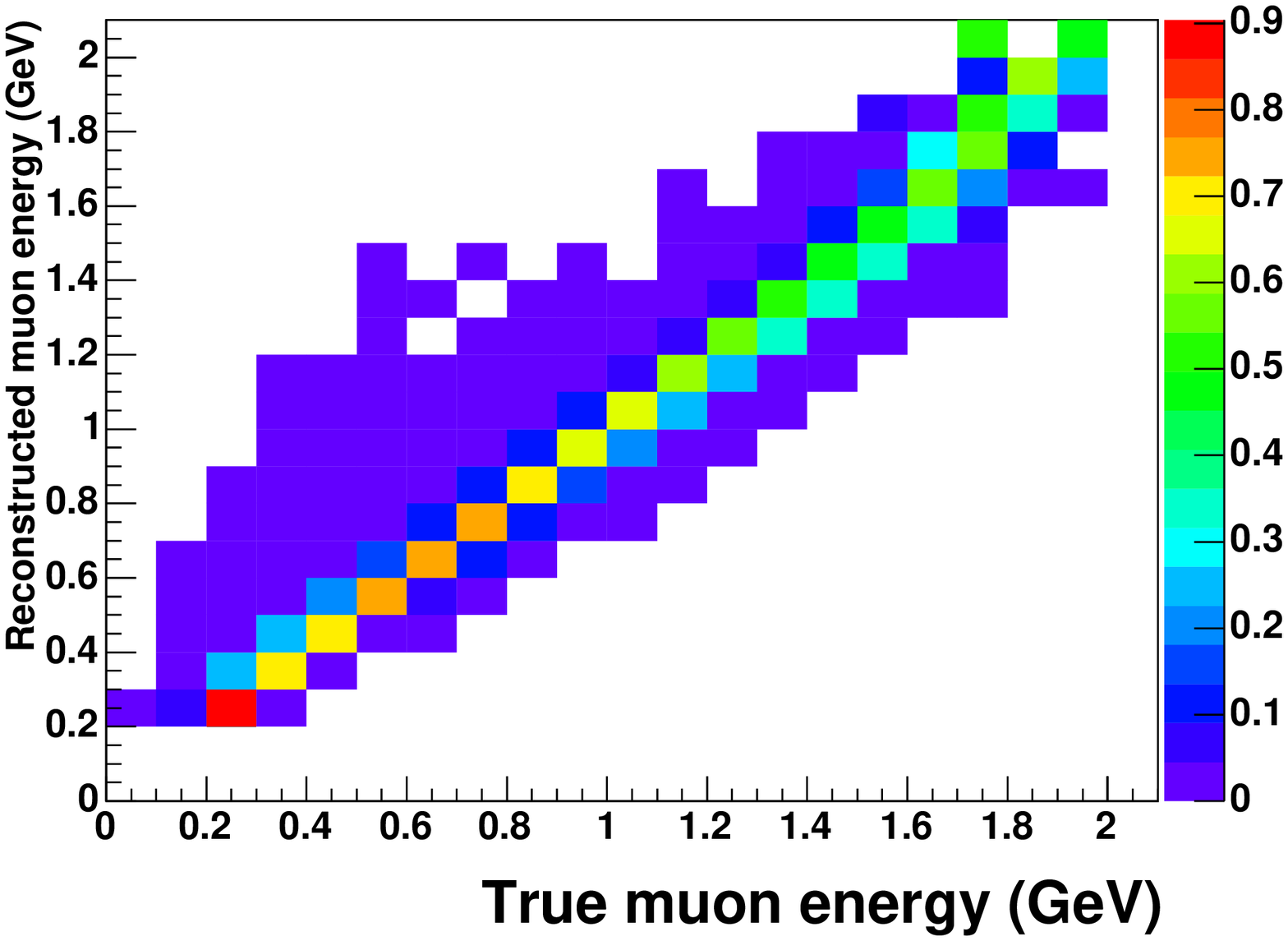,width=2.5in}
}
\vspace*{8pt}
\caption{\label{fig:xsec_bkgd}
The left plot shows the measured muon energy spectrum along with the total predicted background
including the subset of irreducible events. 
The right plot shows the unsmearing matrix for muon kinetic energy. 
%As you see, the distribution is symmetric. 
Note this matrix depends on the detector model; however in this case
the dependency is weak and the resulting uncertainty is negligible compared to other 
errors. 
}
\end{figure}
 
The predicted background is removed from the data (Fig.~\ref{fig:xsec_bkgd}, left).
Since the process does not depend on the prediction of the signal channel, 
the background subtraction method, $d_j-b_j$, is recommended because 
it is less model dependent than purity correction method, 
$d_j\times\frac{s_j}{s_j+b_j}$, where predicted signal and the background in the $j$th bin are used to 
calculate the purity of the signal. 
Since we are measuring the CCQE cross section to allow to study theoretical CCQE models, 
the cross section model for the same process should not enter the formation of the 
physics sample. 
%%As we see, this is violated in our unfolding method.

There is one caveat of background subtraction method. 
In many cases the background prediction is given as a fraction of events. 
To reliably subtract the background, it is important to know 
the absolute scale of the backgrounds.  This is typically obtained from side band studies and this is 
also the case for MiniBooNE. 
For MiniBooNE cross section analyses, 
as many irreducible backgrounds as possible are directly measured in side band samples prior
to subtraction\cite{Teppei_MBSB}.

%Subsequent to background subtraction, we obtain the CCQE sample.  However, 
%its distribution is also unfolded assuming CCQE interaction. 
An alternative cross-section configuration involves to background subtraction at all.  This can 
be used to obtain differential CCQE-like cross sections. 
Such CCQE-like data can allow studies of pion absorption channels that are important 
in the irreducible prediction\cite{Mosel_enu}. 

\subsection{Unsmearing, $U_{ij}$}

The measurement is smeared due to various detector-related effects, and 
event selection unavoidably biases the data sample. 
Unfolding is the process to remove these biases. 
Correcting the data for these processes can be separated through the use of the unsmearing matrix $U_{ij}$, and 
accounting for the detection efficiency $\frac{1}{\ep_i}$. 
For unsmearing, the iterative Bayesian method\cite{DAgostini} 
is popular for all MiniBooNE analyses~\cite{Teppei_MBSB}, 
including $\numu$($\numubar$) CCQE analyses.  This is primarily because the performance is guaranteed for any shapes and any number of bins in the distributions. 
This method is a Bayesian approach, which means measured data distribution is transformed to the unsmeared  
true distribution given by the simulation (Fig.~\ref{fig:xsec_bkgd}, right). 
Therefore, the process depends on the model used to produce this unsmearing matrix. 
This introduces strong model dependencies for inferred variables, 
such as neutrino energy or 4-momentum transfer, however, 
biases to measured variables, such as muon energy and angle, are small. 
The unsmearing process is repeated with different unsmearing matrices 
based on different prior distributions (different CCQE models), 
to estimate the systematic error. 
In this sense, measurements using this technique can never be completely free from CCQE model dependency.  Fortunately
in this case the systematic error from a conservatively large range of prior distributions results in an uncertainty negligible compared to other, better understood errors. 
Different unsmearing techniques are studied in other analyses\cite{Colin_thesis}. 
Unsmearing is a deep subject and future experiments are encouraged to explore further. 
 
%%In this sense, the measurement is not completely free from CCQE model dependency.  However, the shift in
%%the calculated cross sections due to unfolding the
%%data using  conservative range of CCQE model parameters is negligible compared to the flux uncertainties. 

\subsection{Efficiency correction, $\ep_i$}

\begin{figure}[tbp]
\centerline{
\psfig{file=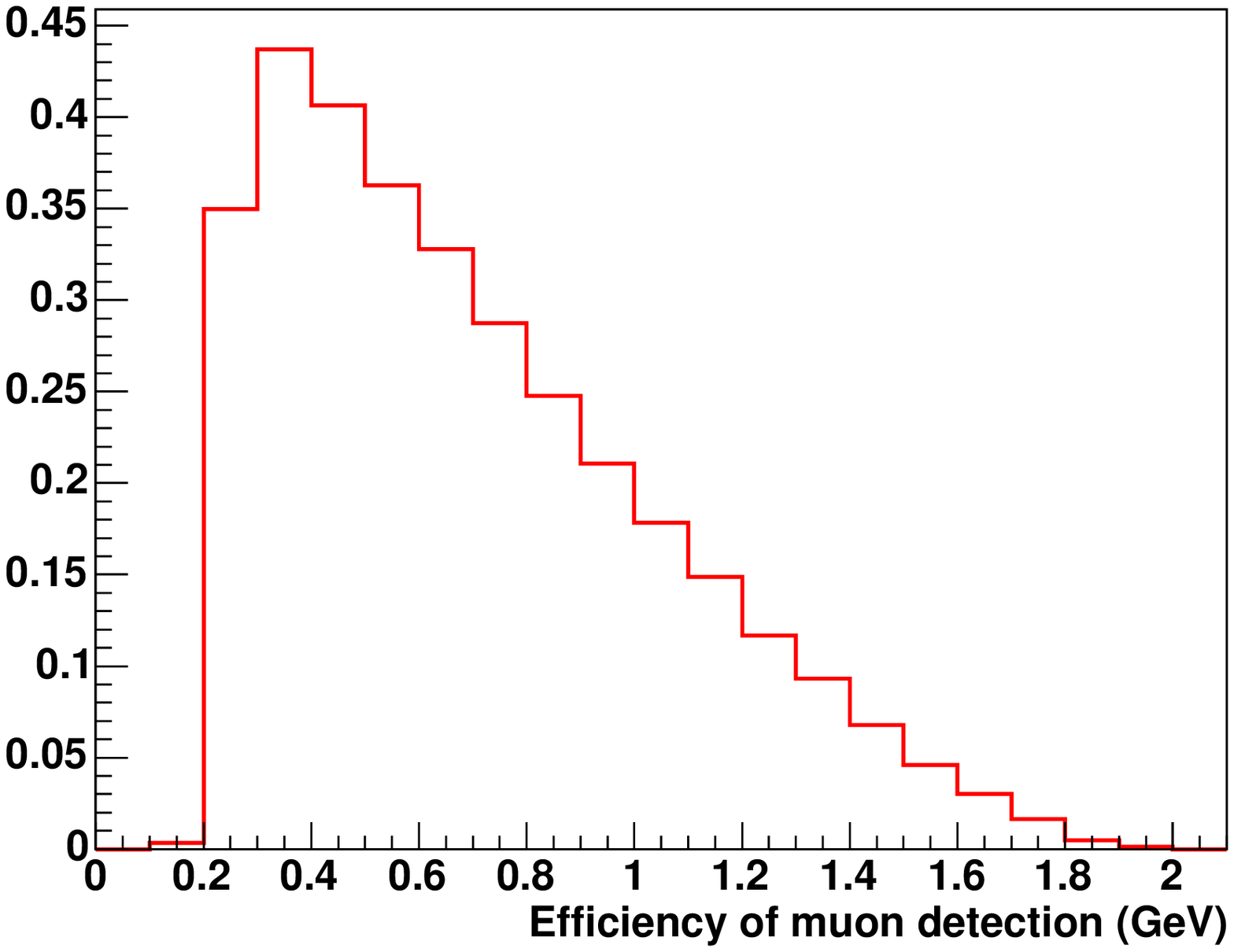,width=2.5in}
\psfig{file=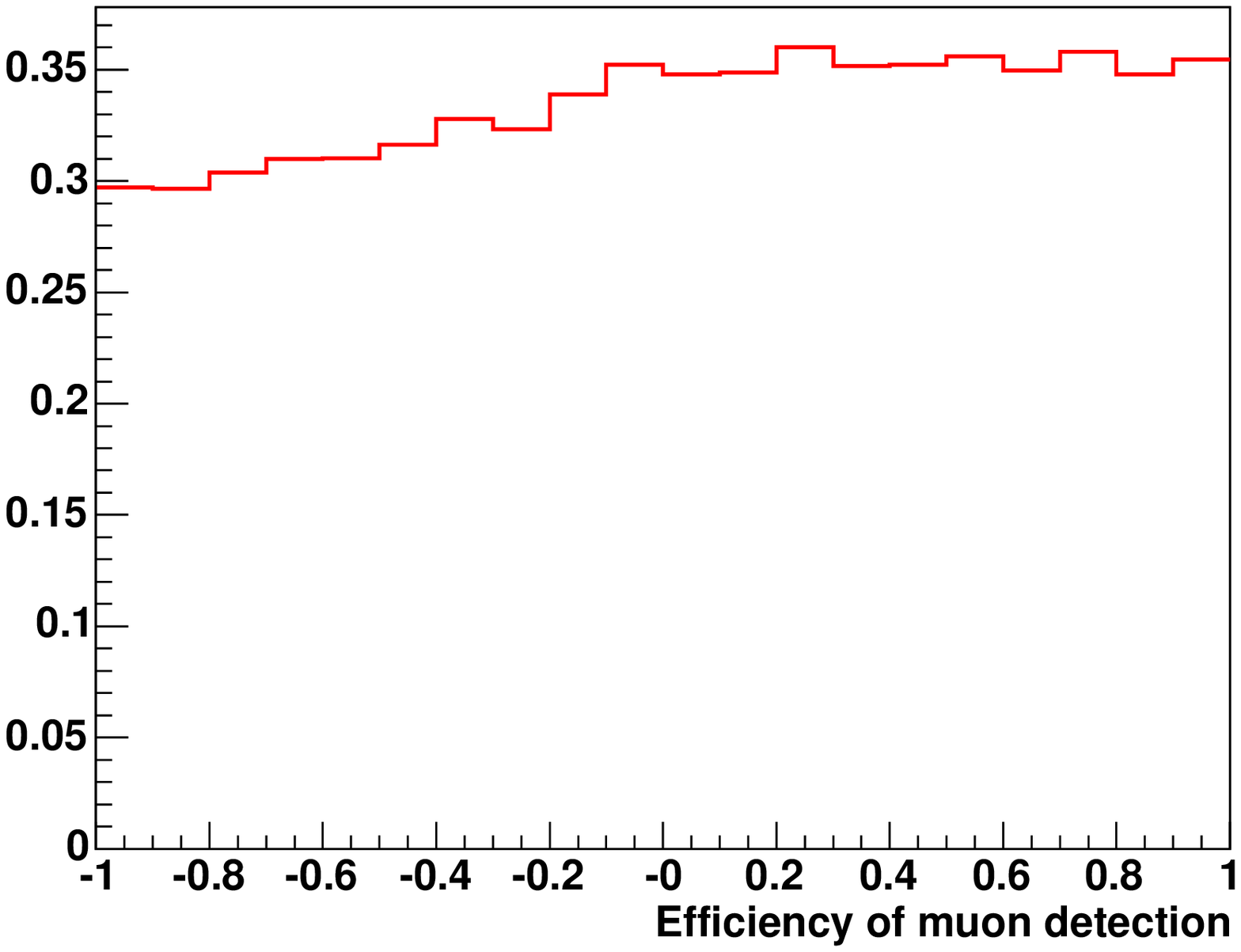,width=2.5in}
}
\vspace*{8pt}
\caption{\label{fig:xsec_effi}
These plots show the detection efficiency of muons from CCQE events in the MiniBooNE detector.  
In the left plot, the efficiency of forward going muons ($0.9<cos\th_\mu<1.0$) 
is shown with function of muon kinetic energy. 
In the right plot, the efficiency of low energy muons ($200<T_\mu~(\uMeV)<300$) 
is shown with function of muon scattering angle.   
}
\end{figure}

The efficiency correction is applied to recovered the distribution 
``what if there are no cuts?'' utilising efficiency, $\ep_i=\frac{N^{after~cut}_i}{N^{before~cut}_i }$, 
which is the ratio of signal events in the simulation after all cuts ($N^{after~cut}_i$) to before cuts ($N^{before~cut}_i$). 
Fig.~\ref{fig:xsec_effi} show the efficiency of muons from CCQE interactions. 
From the left plot, one can see the efficiency drops monotonically with increasing the muon energy, 
simply because the detector cannot contain higher energy particles. 
On the other hand, right plot shows the efficiency is rather constant with muon scattering angle, 
because the rotationally symmetric MiniBooNE detector has uniform efficiency over scattering angles. 
Note there is some loss of efficiency at backward going muons due to the presence of high energy protons. 
 
%Since many neutrino experiments are designed to measure oscillation phenomena,
One must exercise caution with the efficiency correction. Data are often used only in limited phase space, 
such as in a small fiducial volume, 
narrow kinematic space etc. 
Since efficiency correction may recover events outside of that, 
simulation needs to be understood in the phase space in which we do not perform any measurements. 
For MiniBooNE CCQE analyses, we do not measure muons if kinetic energy is below 200~MeV. 
However, muons with energy lower than 200~MeV also contribute to the final sample due to smearing effect (the opposite is also true). 
Therefore it is possible to recover events in the kinematic space we do not perform measurements. 
Since such restoration of data heavily rely on the detector simulation, 
we choose not to report measurements with muon energy lower than 200~MeV. 
Similarly, the signal region is defined as a 550~cm radius sphere, which is smaller than the inner target volume 
(575~cm sphere) to avoid effects not simulated well. 

\subsection{Target number and flux correction, T, $\Phi$}

Finally, important normalization factors needed to obtain the differential
cross section are the target number ($T$) and 
total flux of interaction ($\Ph$). 
%%The target number is the full volume of interactions are assumed to happen, 
%%and its uncertainty is basically from the uncertainty of the fiducial volume. 
The total flux is the number of neutrino involved in the measurement of interest. 
Therefore, each measurement has a different number of 
total flux due to the presence of interaction thresholds. 
This concept is important especially to estimate the error. 
In general, modern wideband beam has a large error at the lowest neutrino flux distribution. 
However, such low energy neutrinos will not contribute to certain type of interactions 
such pion production channels. 
Therefore, taking account of threshold effects of the interaction in general 
give smaller flux uncertainties.
 
\subsection{Systematic error}

%The differential cross section measurement has four major systematic errors, 
%signal model error, background model error, flux model error, and detector model error.  
%Table~\ref{tab:sys} show the list of systematic errors contributed on each processes. 

The details of systematic errors can be found elsewhere\cite{MB_CCQE,MB_ANTICCQE}, 
but flux uncertainties dominate the normalization error 
($\sim$9\% in $\numu$CCQE, $\sim$10\% in $\numubar$CCQE cross section)
for CCQE interactions in MiniBooNE. 
%Detector uncertainties large shape error contribution 
%both low and high energy region. 

Here we briefly discuss the propagation of errors. 
With the background subtraction method and iterative Bayesian unfolding method, 
the systematic error of the signal channel contributes mainly through the unsmearing matrix. 
In fact, signal channel MC is used to calculate efficiency, 
so there is some contributions on efficiency, too. 
However, the efficiency is defined by the ratio of signal MC after cuts to before cuts, 
and many systematics cancel in this ratio.  
The background model affects any background removing process, and it is 
crucial to constrain this prediction with external data or side band measurements
as much as possible. 
%The detector effect mainly affect unfolding, both unsmearing and efficiency correction. 
%%This is the most important effect to remove, and an experiment should be very careful and confident. 
The neutrino flux error is the biggest normalization error, and mainly comes from the flux factor $\Ph$, 
but also can modify the scale of the background prediction.  This provides another motivation for {\it in situ}
background measurements.
%This is another motivation for {\it in situ} background measurements,
%because 
%%because then you can apply the constraint from the measurement. 

%%\begin{table}
%%\tbl{List of systematic errors in each cross section calculation process.}
%%{\begin{tabular}{@{}lcclc@{}} \toprule
%%\hline
%%Systematics type&background model &Unsmearing matrix &Efficiency &Total flux\\
%%\hline
%%Signal model    & none            & big              & small     & none \\
%%Background model& big             & none             & none      & none \\
%%Detector effect & small           & big              & big       & none \\
%%Neutrino flux   & big             & small            & small     & big  \\
%%\hline
%%\end{tabular}}
%%\label{tab:sys}
%%\end{table}

\section{CCQE in $\nu$-mode\label{sec:numuccqe}}

\subsection{CCQE interaction model data driven correction\label{subsec:mafit}}

%Table~\ref{tab:sel} lists the requirements of the physics sample .  
%Broadly, the selection is entirely based on simple kinematic observations of 
%the Cherenkov light from prompt muon and it's decay electron to ensure the CCQE events studied are well-reconstructed, 
%while also optimising the detection efficiency and accepting background interactions as low as feasible.

%%\begin{table}
%%\tbl{List of cuts for CCQE selection.}
%%{\begin{tabular}{@{}lcclc@{}} \toprule
%%\hline
%%Cut Description  \\
%%\hline
%%Two timing clusters \\
%%Veto hits \textless\, 6  \\
%%First timing cluster in beam window  \\
%%$\tmu$ \textgreater\, 200 MeV   \\
%%Reconstructed radius within 500 cm \\
%%reconstructed $\mu$-$e$ dist. \textgreater\, 500 cm/GeV $\times\,\tmu$ - 100 cm  \\
%%reconstructed $\mu$-$e$ dist. \textgreater\, 100 cm \\
%%ln ($\mu$ / $e$) \textgreater\, 0  \\
%%\hline
%%\end{tabular}}
%%\label{tab:sel}
%%\end{table}

\begin{figure}[tbp]
\centerline{
\psfig{file=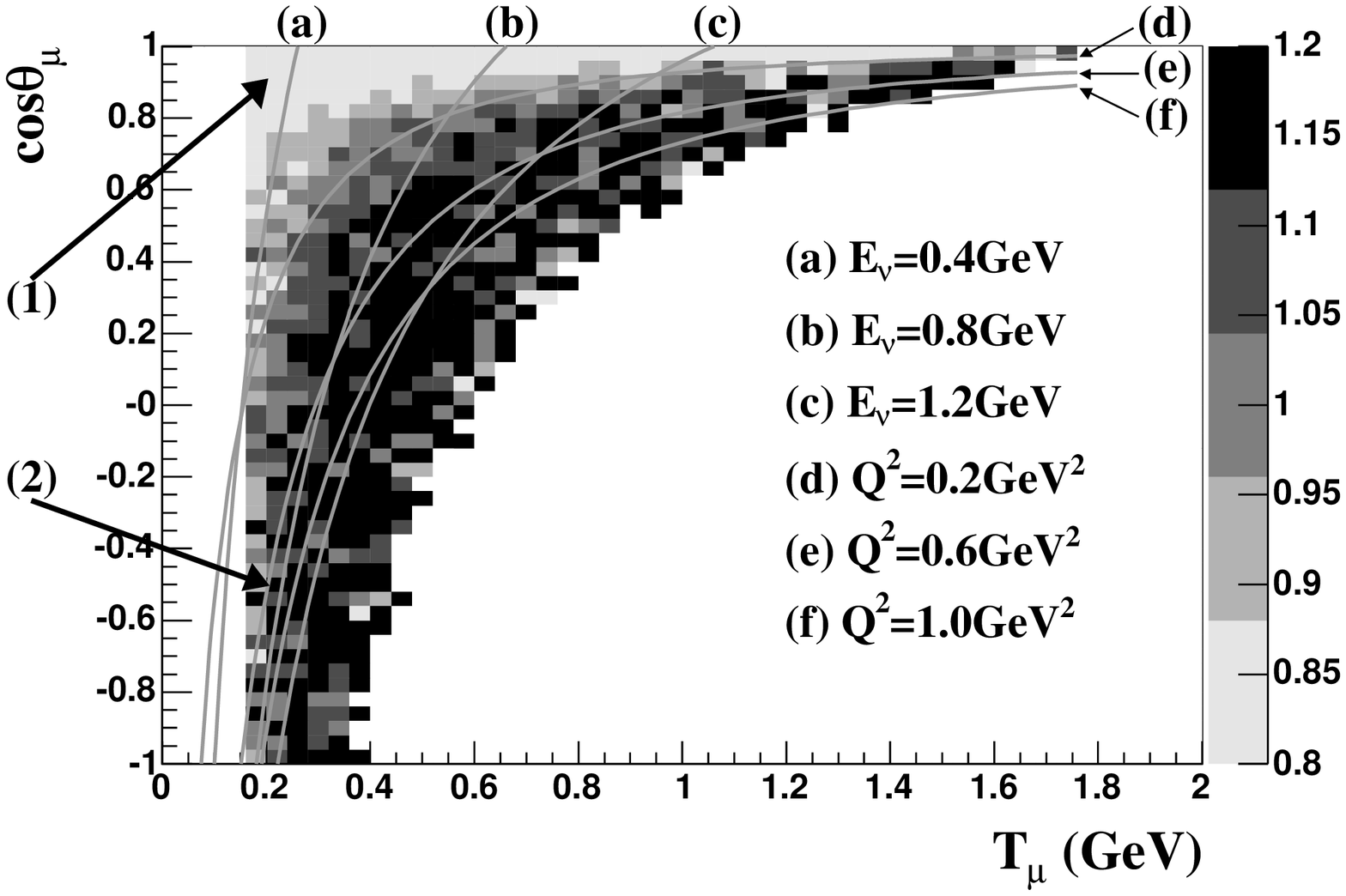,width=2.3in}
\psfig{file=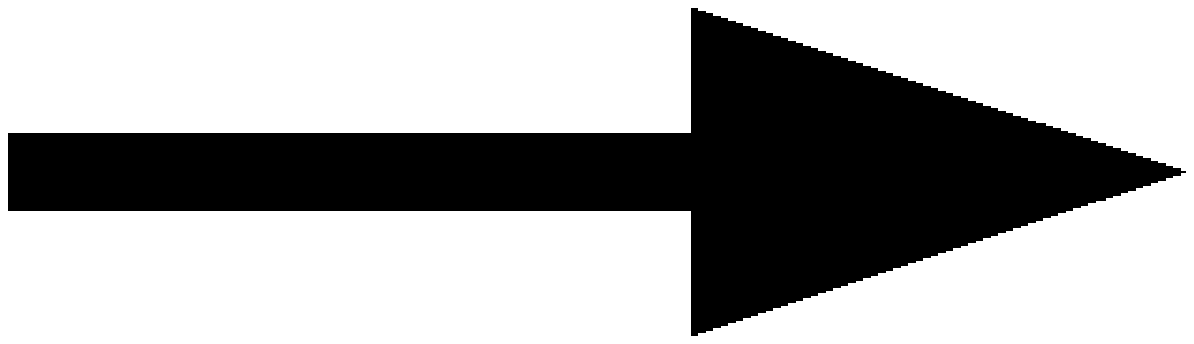,width=0.4in}
\psfig{file=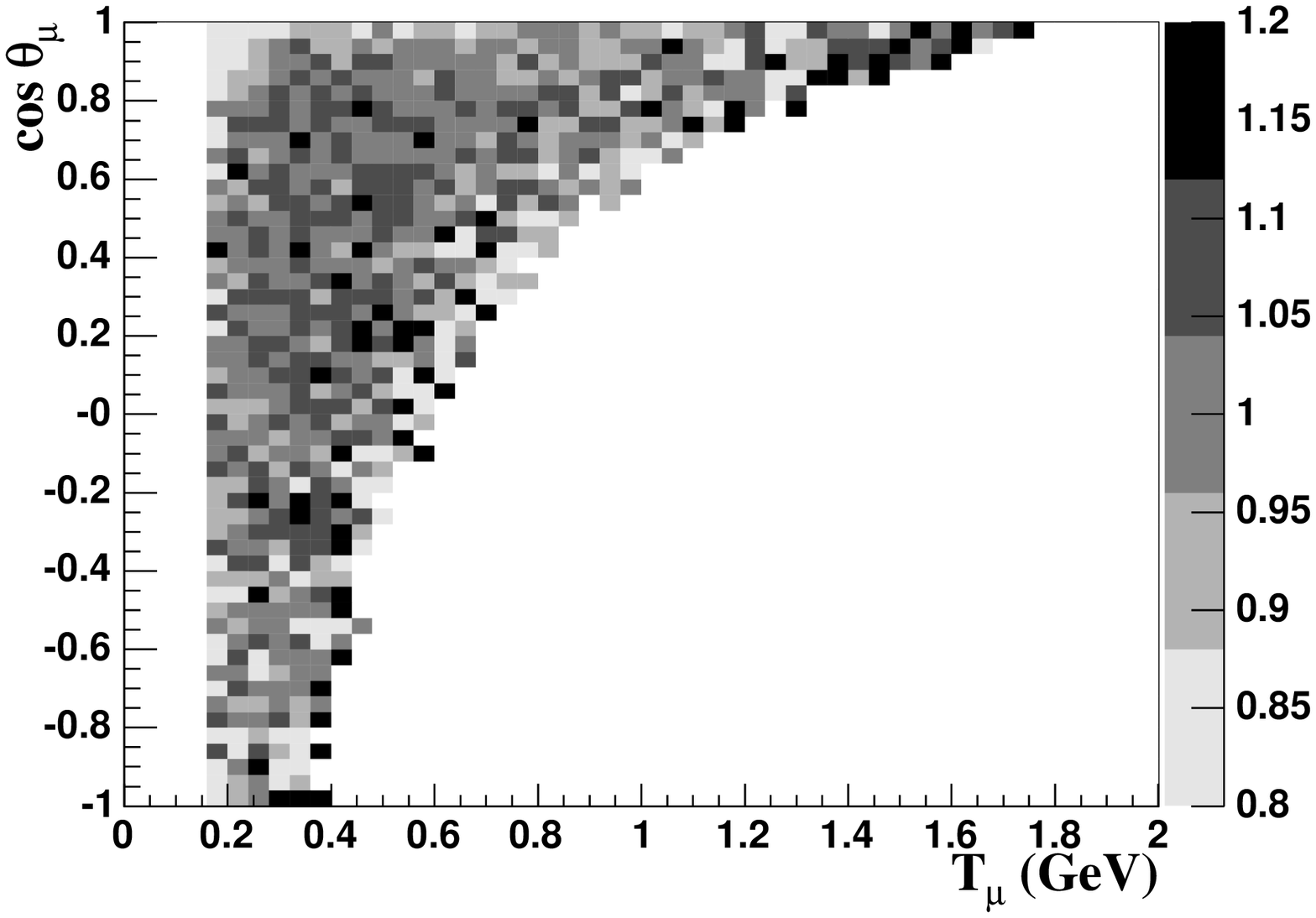,width=2.3in}
}
\vspace*{8pt}
\caption{\label{fig:CCQEPRL} Data-simulation ratio of measured CCQE events. 
Note simulation is normalized to data statistics.}
\end{figure}

MiniBooNE was the first experiment to collect statistics sufficient to make
a detailed comparison between the observed and predicted muon kinematics for 
all possible production angles~\cite{MB_CCQEPRL}.  
Presented as their ratio, the measured distribution is compared with MC simulation in Figure~\ref{fig:CCQEPRL}  (left). 
There are clear discrepancies at two regions indicated by arrows. 
Top left corner (region 1) shows overestimation of the simulation, 
and middle black band (region 2) indicate simulation underestimates the data. 
The questions is, what is wrong in the simulation? 
Since data is a convolution of neutrino flux, the neutrino cross section, and the detection efficiency, 
data-simulation discrepancy can be any of mis-modelling of them. 
However, this problem can be disentangled by following way. 

Here, neutrino flux $\Ph$ is a function of neutrino energy, 
and cross section model $\si$ is a function of $Q^2$, 
and as we see detector efficiency is function of muon energy $T_\mu$. 
If we put all knowledge together, 

\beq
rate\propto\int\Ph(E_\nu)\times\si(Q^2)\times\ep(T_\mu)
\label{eq:xsdep}
\eeq

Now, the Fig.~\ref{fig:CCQEPRL}, left, has 6 curves, 
representing reconstructed neutrino energy $E^{QE}_\nu$ (a, b, and c) 
and 4-momentum transfer $Q^2_{QE}$ (d, e, and f). 
The data-MC ratio plot clearly shows discrepancy along the reconstructed $Q^2$ lines, 
meaning simulation is wrong with function of $Q^2$, 
namely we have some problem with neutrino cross section model. 
This is a strong motivation to fix the CCQE cross section model.  
%which is also used for the oscillation search. 
Two effective parameters are introduced in a CCQE cross section model, 
and as you see, simulation agree well in entire kinematic space after 
the correction (Figure~\ref{fig:CCQEPRL}, right). 
This parametrization was motivated purely by the practical need to have an adequate description of
$\numu$ CCQE events before the ``box" could be opened on the $\nue$ CCQE candidates.  
%The {\it a priori} disagreement with the MC prediction was suspected to be due to the naive modelling of nuclear physics.

\subsection{Results}

Now, using the formula (Eq.~\ref{eq:dd}), 
we are ready to calculate the CCQE flux-integrated double differential 
cross section: 

\beq
\frac{d^2\sigma}{d\tmu\,d\left(\uz\right)}_i = \frac{\sum_j U_{ij}\left(d_j - b_j\right)}{\Delta\tmu\,\Delta\uz \,\epsilon_i \,\Phi \,T}, 
\label{eq:dd}
\eeq

\begin{figure}[tbp]
\centerline{
\psfig{file=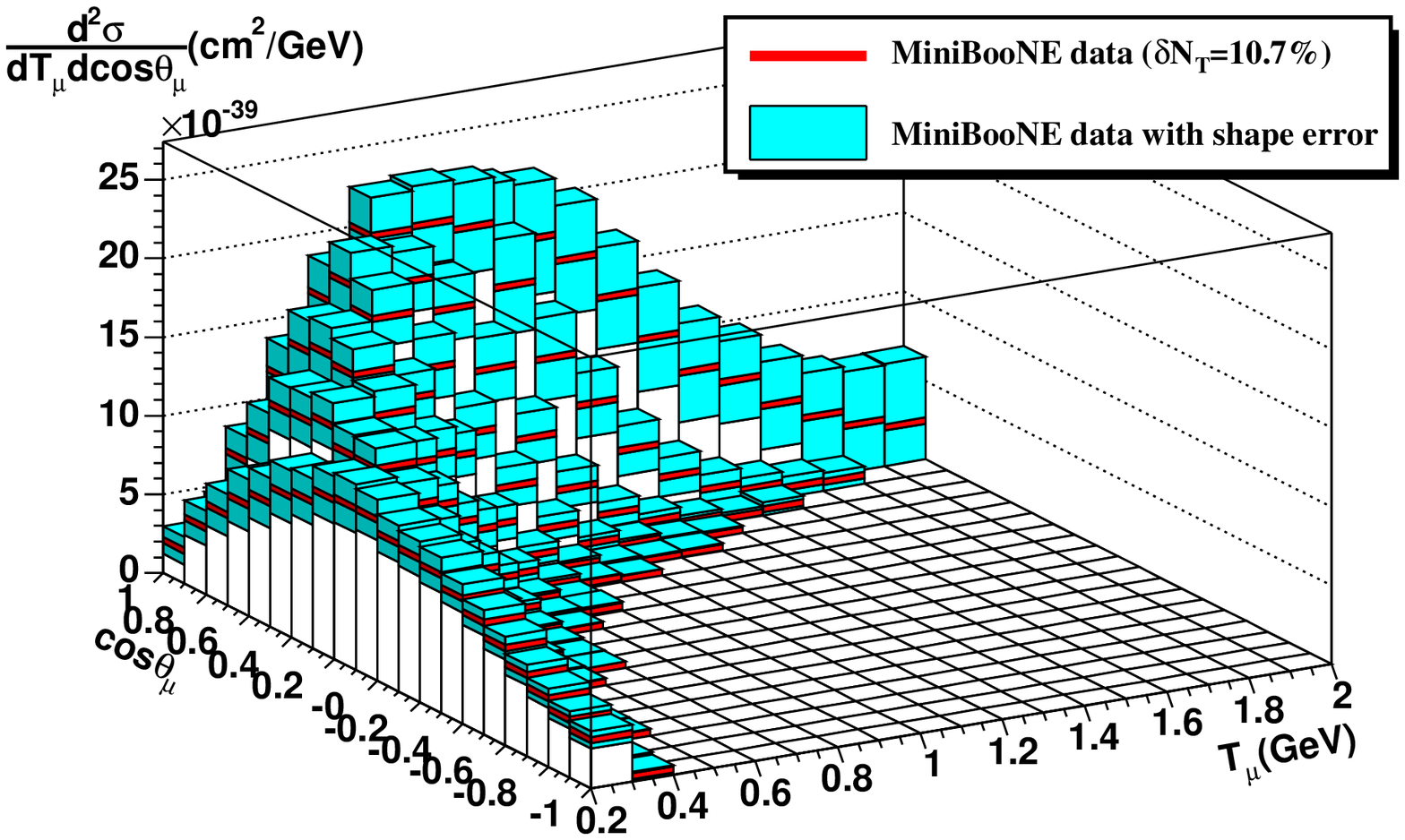,width=2.5in}
\psfig{file=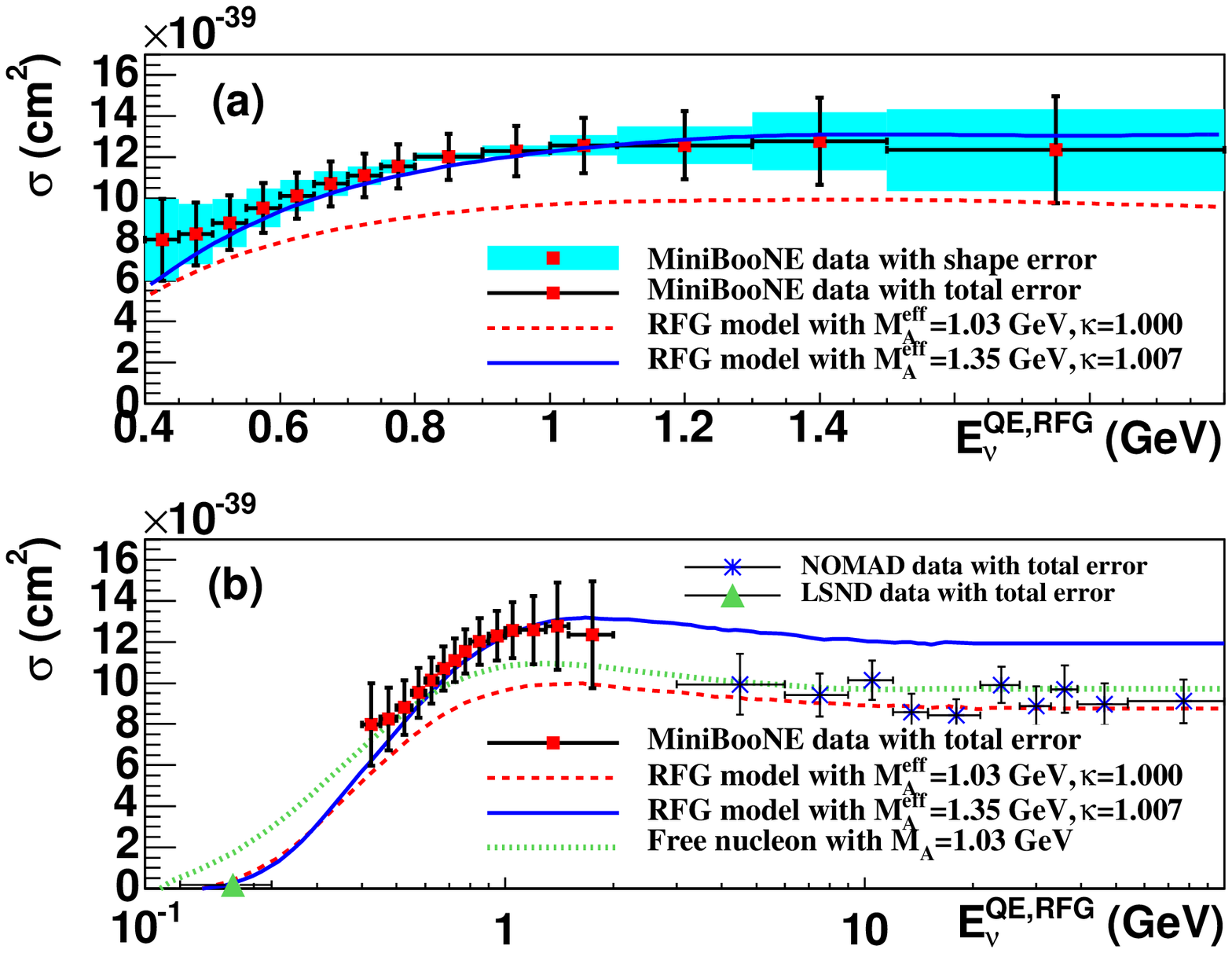,width=2.5in}
}
\vspace*{8pt}
\caption{\label{fig:numuxsec} 
$\numu$CCQE cross section results. 
Left is the flux-integrated double differential cross section, 
and the right is the flux-unfolded total cross section.}
\end{figure}

This data has the highest amount of information possibly measured in this topology, 
with least model dependencies.  
Figure~\ref{fig:numuxsec} shows the result\cite{MB_CCQE}. 
The left plot shows the double differential cross section. 
An attractive feature of this result is
that any theorists may simulate this by convoluting their models with given BNB flux 
without need of simulation software used by the experiment. 
The right plot shows the flux-unfolded total cross section. 
Although the total cross section can nicely compare the result with other experiments, 
reconstructing neutrino energy in MiniBooNE energy region is highly model 
dependent\cite{Mosel_enu,Martini_enu1,Martini_enu2,Nieves_enu}, 
and the comparison with theoretical models and other experiments' results need care. 
These plots show the clear disagreement with the relativistic Fermi gas (RFG) model. 
The normalization of the measured cross section is much higher than 
the (RFG) model with world-averaged parameters ($M_A=1.03$~GeV). 
%Here, we added back irreducible backgrounds, so the result can be understood CCQE-like cross section.  
%The result is the absolute value, and it shows large normalisation mismatching from existing CCQE cross section models. 

In summary, MiniBooNE CCQE data revealed three mysteries:
\begin{itemize}
\item low $Q^2$ suppression
\item high $Q^2$ enhancement
\item large normalisation
\end{itemize}

It is still under the debate, but the community slowly understands the underlying
details of these results. 
First, it was shown that the violation of 
the impulse approximation (IA) is severe at low $Q^2$ region\cite{Butkevich_rfg,Ankowski}. 
Many interaction models are based on the IA, 
where a probe particle is assumed to interact with single nucleon. 
This approximation is broken at low momentum transfer ($<300~$~MeV/c), 
where space resolution of probe particle is larger than inter nucleon space. 
Some calculation, such as random phase approximation (RPA), 
properly takes this low energy physics into account.
For the high $Q^2$ region, 
it was pointed out that interaction involved in two nucleons (two-body current), which is also neglected by IA, 
has an appreciable effect, and the calculation including 
two body currents can reproduce both high $Q^2$ shape and normalisation\footnote{
The contribution of the two-body current in $\sim$1~GeV neutrino interaction is one of the most active 
topic in neutrino interaction physics community. 
We refer the reader to the growing body of literature elsewhere~\cite{Zeller_QEreview,Sobczyk_QEreview}.}. 
Models including RPA correction and two-body currents have been proposed\cite{Martini_mec,Nieves_mec},  
and those models show excellent agreement with the CCQE double differential 
data\cite{Martini_dd,Nieves_dd,Mosel_dd}. 
Therefore these models kill three birds with one stone!

However, this is not the end of the story. 
The large enhancement induced by two-body current is not always observed\cite{Donnelly_mec}, 
and also models with different approaches also agree well with MiniBooNE 
data\cite{Bodek_tem,Sobczyk_tem,Meucci_rgf,Butkevich_dwia}. 
Interestingly, some models have different predictions in anti-neutrino CCQE cross sections. 
Therefore, now we are on the very exciting position; 
our task is to provide the antineutrino CCQE cross section 
to test these models.
 
\section{CCQE in $\nubar$-mode\label{sec:numubarccqe}}

\subsection{Introduction}

Mechanically, the creation of an antineutrino beam is trivial: by reversing 
the polarity of the magnetic focusing horn at the proton collision station, the 
$\numu$ parent $\pip$ particles are now largely deflected from the beam, and the 
$\pim$ are focused toward the detector and decay to create a beam of $\numub$.  
In every such antineutrino-mode setup, some important consequences arise: 

\begin{itemize}%[itemsep=0mm]

\item {Event rates}: before considering horn focusing, the leading-particle 
effect leads to the creation of roughly twice as many $\pip$ as $\pim$.  This naturally
leads to a less intense $\numub$ beam.  Moreover, $\numub$ cross sections with matter are 
generally lower compared to the $\numu$ case.  The size of these effects are a function of
energy, and at MiniBooNE the overall neutrino-mode rate difference is around five times higher
compared to the antineutrino-mode rates.

\item {Beam energy}: due also to the leading particle effect, more energy tends to 
be given to $\pip$'s compared to $\pim$.  At MiniBooNE, the mean beam energy for neutrino 
(antineutrino) mode is $\sim$ 800 (650) MeV.  Among other things, this has implications for neutrino 
oscillation measurements, where some regions of $\Delta m^2$ may be more or less challenging to measure
with $\nub$ compared to $\nu$.

\item {$\numu$ contamination}: at least for non-magnetized detectors, by far the most 
experimentally-challenging feature of the antineutrino-mode beam is the natural $\numu$ 
background.  Since the event rates of $\numu$ far surpass that of $\numub$, if the horn 
focusing system accepts even a small amount of $\pip$ into the beam this may lead to large 
rates of $\numu$ interactions in the detector.

\end{itemize}

%In this way, observations of the neutrino and antineutrino rates in the detector
%while varying the horn current polarity and strength allows for tests of various regions of 
%meson production phase to be tested (whether constrained or unconstrained by external data). 

Though the CCQE reconstruction and sample selection were well defined relatively early on in 
MiniBooNE's antineutrino mode data run, the final observation above had to be carefully dealt 
with before a $\numub$ CCQE cross section could be extracted with any reasonable precision.

\subsection{Data driven correction on forward going pion production\label{subsec:forwardpi}}

The analyses described here are available in detail elsewhere\cite{MB_ANTICCQE,MB_ws}.  Here we 
focus on the highlights and the pieces most historically relevant, including applications
to neutrino oscillation experiments looking to measure CP violation without a magnetic field.

\begin{figure}[tbp]$
\begin{array}{cc}
\includegraphics[scale=0.33]{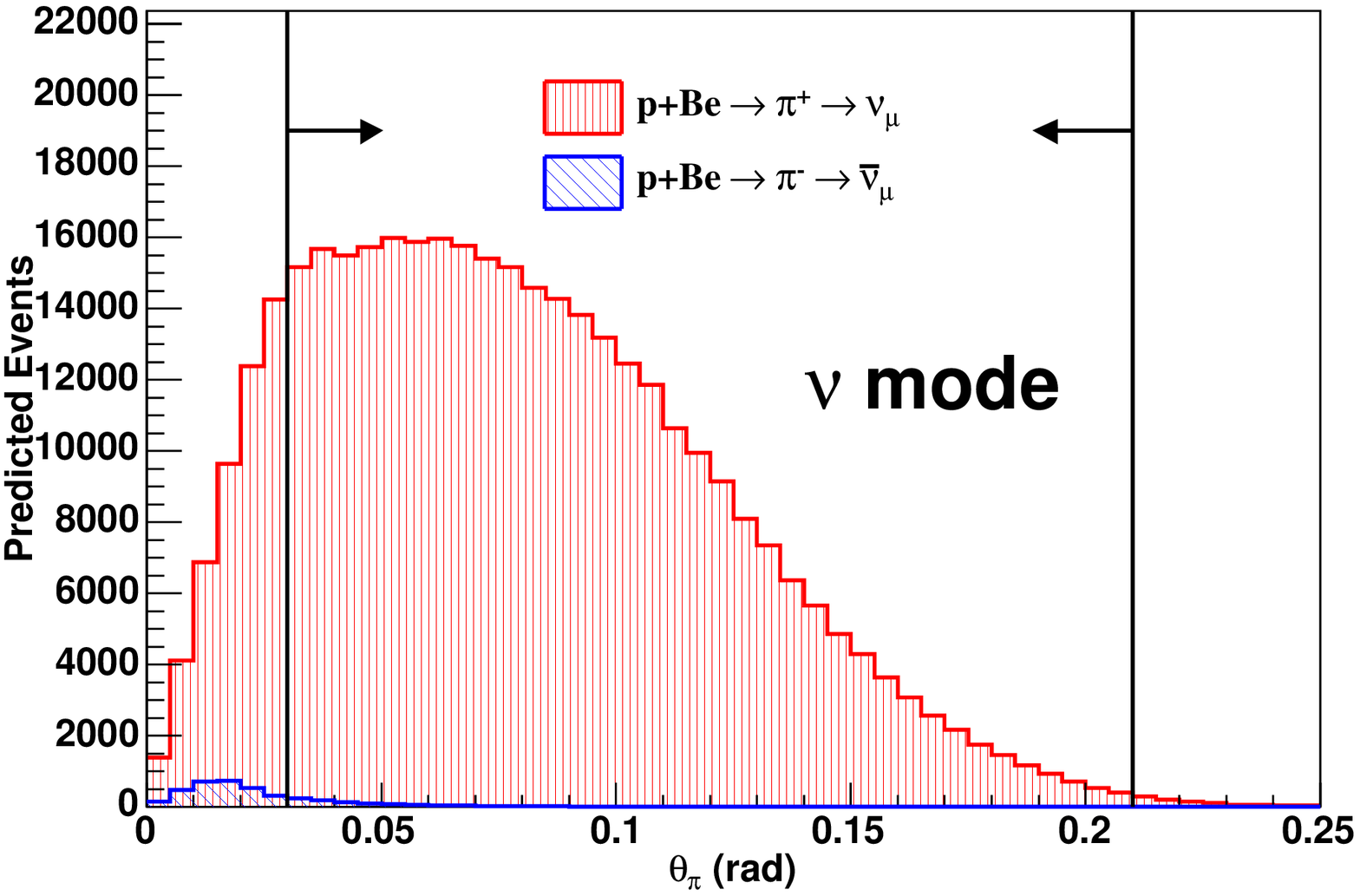} &
\includegraphics[scale=0.33]{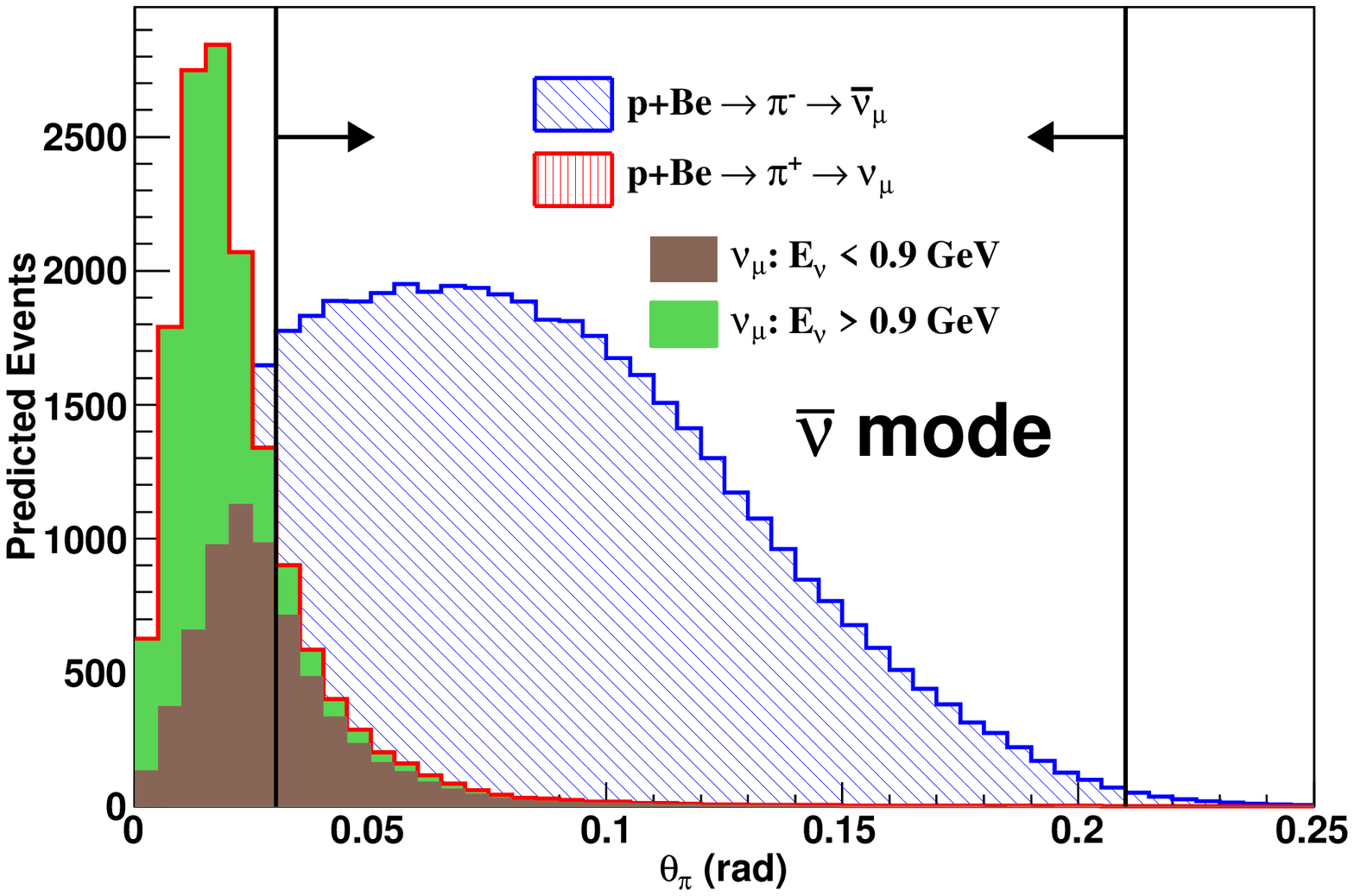} \\
\end{array}$
\vspace*{8pt}
\caption{\label{fig:piProd} Pion production angular distributions with respect to the incident proton beam ($\theta_{\pi}$) 
producing $\numu$ and $\numub$ in neutrino (left) and antineutrino (right) modes.  Arrows indicate the region constrained by HARP data.}
\end{figure}

The largest background to the $\numub$ CCQE sample comes from $\numu$ CCQE interactions. 
As shown by Fig.~\ref{fig:piProd}, 
these events are accepted into the antineutrino-mode beam through high-energy and 
forward-going $\pip$.

A few more remarks regarding Fig.~\ref{fig:piProd}:

\begin{itemize}%[itemsep=0mm]
\item the antineutrino contribution to the neutrino-mode data is negligible in comparison to the converse.  
This is due to a convolution of flux and cross-section effects that simultaneously serve to enhance 
the neutrino component while the antineutrino contribution is suppressed. 
%%\bf{TK: this is mentioned}
%%the leading-particle effect at the beryllium target 
%%(the $p + \textrm{Be}$ initial state has a net positive charge) naturally leads to the creation of roughly twice as many $\pip$ as $\pim$, 
%%and neutrino cross sections are typically around three times as large as antineutrino cross sections around 1~GeV.
\item as seen in the antineutrino-mode distribution, high-energy $\numu$'s are 
strongly correlated with the decay of $\pip$ created at very small opening angles.  
This indicates their flux is more poorly constrained by the HARP data compared to lower-energy $\numu$'s.  
So, not only is the overall $\numu$ flux in antineutrino mode highly uncertain, 
the accuracy of the extrapolated $\numu$ flux prediction may be a function of neutrino energy.
\end{itemize}

Each of these issues will be encountered by future neutrino oscillation experiments that will observe antineutrino-mode
beams though non-magnetized detectors.  These complications have never before been dealt with, and so
 MiniBooNE's approach was an aggressive campaign to measure the $\numu$ contamination with as many complementary analyses 
as possible.  Three analyses offered comparable uncertainties and low levels of systematic correlations:

\begin{enumerate}%[itemsep=0mm]
\item {\bf $\pi^+$ decay in $\numu$CC1$\pi^+$ and $\pi^-$ capture in $\numubar$CC1$\pi^-$}:
The $\pim$ produced in $\numub$ CC1$\pi$ interactions ($\numub+N \to \mup+N+\pim$) is captured on carbon nuclei. 
A key difference is the rate for stopped-$\pim$ nuclear capture is nearly 100\%, 
and so the simple sample selection of $1\mu+2e$ 
($\mu$ and $\pi^+$ decay chains make 2 electrons, total 3 subevents) 
yields a highly-pure sample of $\numu$ events from which to measure 
their overall contribution to the antineutrino beam.
\item {\bf CCQE angular distribution}:
In general, $\numub$ CC interactions feature much lower momentum transfer compared to $\numu$.  
This leads to a kinematic restrictions for $\numub$ compared to $\numu$, 
and in particular backwards-produced $\mup$ are highly suppressed.  
Therefore, the angular distribution may be used to separate the contributions from $\numu$ and $\numub$. 
It should be noted that this technique depends on detailed knowledge of the $\numub$ cross sections. 
Since we do not know these {\it a priori}, this result is not used to extract the $\numub$ CCQE cross sections. 
However, as experimental and theoretical knowledge of these interactions are rapidly improving, this could be a powerful analysis in the future. 
\item {\bf $\mum$ capture}:
Nuclear capture of $\mum$ near carbon nuclei ($\sim$8\%) affords a statistical asymmetry between $\numu$ and $\numub$ CCQE events.  
By simultaneously analyzing samples of $1\mu$ (1 subevent) and $1\mu$+1e (2 subevents) sample, 
appreciable sensitivity is gained to the relative of $\mup$ and $\mum$ present in each sample. 
Analyses of this kind could be hugely beneficial to next-generation experiments planning to use liquid argon as the detection medium~\cite{snowmass_nu}, 
where the $\mum$ nuclear capture rate on argon is near 70\%.  For such a high rate and based on the presence (or lack thereof) of a decay electron, one could nearly distinguish between $\numu$ and $\numub$ interactions on an event-by-event basis without a magnetic field!
\end{enumerate}

\begin{figure}[tbp]
\centerline{
\psfig{file=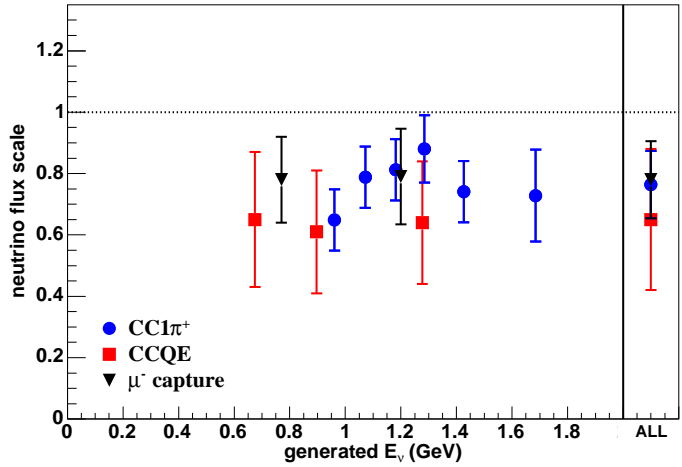,width=4.0in}
}
\vspace*{8pt}
\caption{\label{fig:wsSumm} Summary of the results from three techniques used to measure the $\numu$ flux in the anti-neutrino mode beam.}
\end{figure}

Presented as a constraint relative to the nominal prediction extrapolated from HARP data, 
the final results~\cite{MB_ws} from these analyses are presented in Fig.~\ref{fig:wsSumm}.

With these constraints, the uncertainty on $\numub$ flux from 
the HARP $\pim$ production data forms the largest contribution to the overall $\numub$ CCQE data sample.  
This is effectively an irreducible uncertainty to MiniBooNE, and so we turn our attention to the results.

\subsection{Results}

\begin{figure}[tbp]
%%\centerline{
\psfig{file=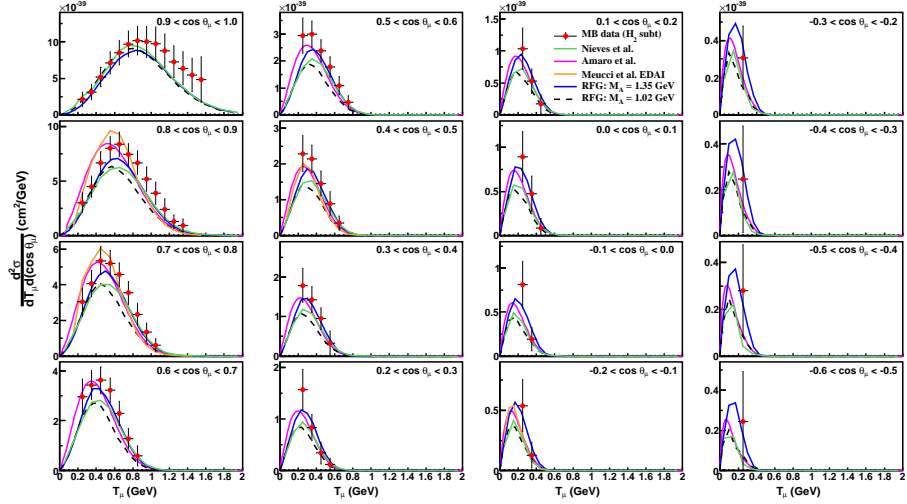,width=6in}
%%}
\vspace*{8pt}
\caption{\label{fig:numubarxsec} 
The $\numub$ CCQE flux-integrated double-differential cross section compared to various predictions.
}
\end{figure}
 
Fig.~\ref{fig:numubarxsec} shows 
a comparison of the $\numub$ CCQE data with the various predictions from 
Nieves {\it et al.}\cite{Nieves_anti}, 
Amaro {\it et al.}\cite{Donnelly_anti}, 
Meucci {\it et al.}\cite{Meucci_anti}, 
and RFG models with different parameter choices. 
The RFG model with world-averaged parameter ($M_A=1.02$~GeV) 
reveal discrepancies found also in the behaviour of $\numu$ CCQE data. 
Namely, a suppression of low-momentum transfer events, 
more cross-section strength for relatively backward-scattered muons, 
and a normalization excess of $\sim$~20\%. 

So what does this mean?  
Interesting to note, the RFG parametrized with a high axial mass ($M_A = 1.35$~GeV) 
seems to do remarkably well describing the antineutrino data.  
While the nuclear model assumed in the RFG is rather naive, 
apparently the interplay between axial and vector pieces in this model is grossly correct.  
This is at least true at the energies probed by MiniBooNE ($E_\nu\sim700$~MeV), 
however, recent data from MINER$\nu$A suggest that this breaks down 
at higher energies ($E_\nu \in 1.5-10$GeV)~\cite{minervaNu,minervaNub}. 

\section{Combined measurements of $\numu$ and $\numub$ CCQE\label{sec:combined}}

\begin{figure}[tbp]$
\begin{array}{cc}
\includegraphics[scale=0.33]{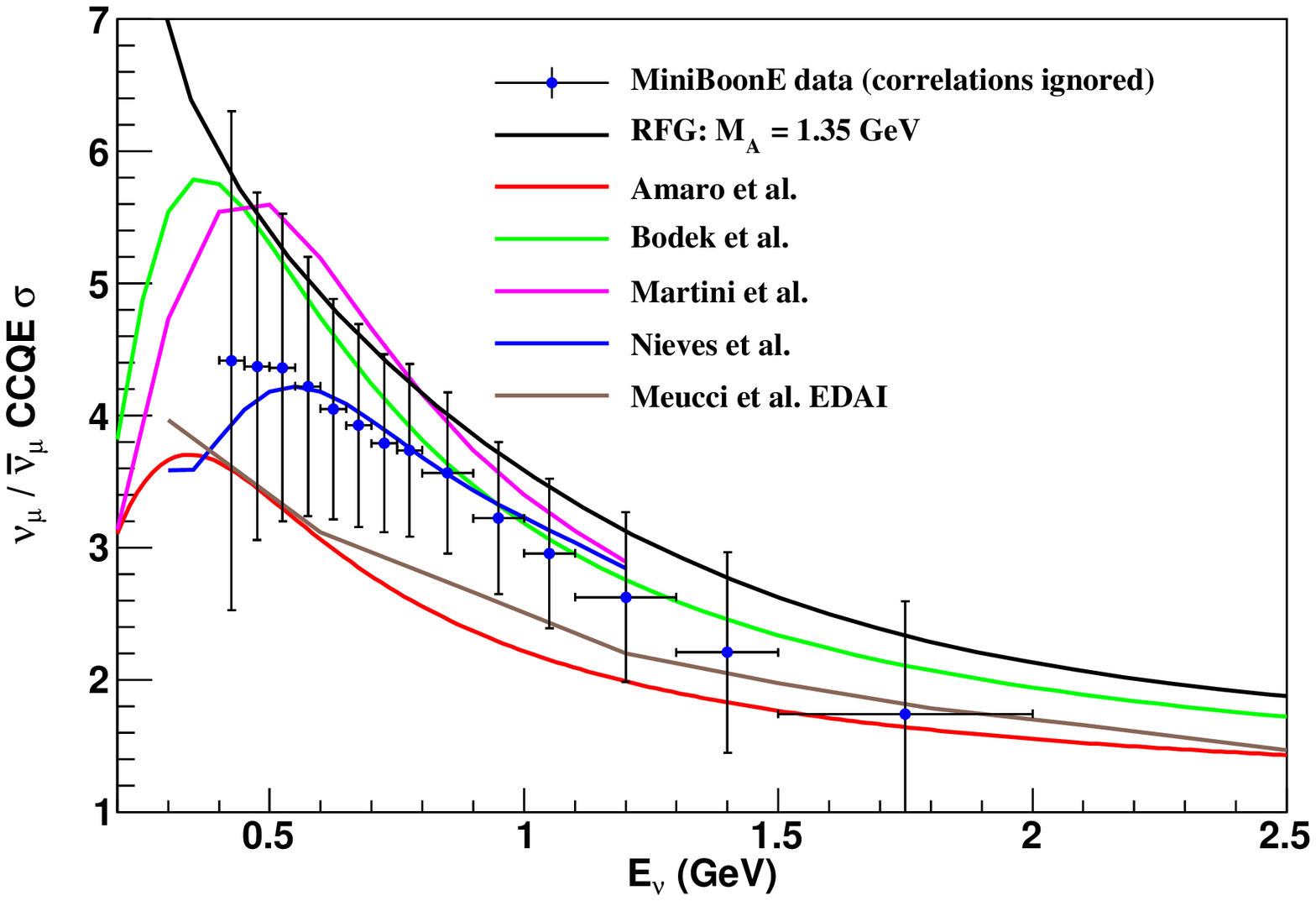} &
\includegraphics[scale=0.33]{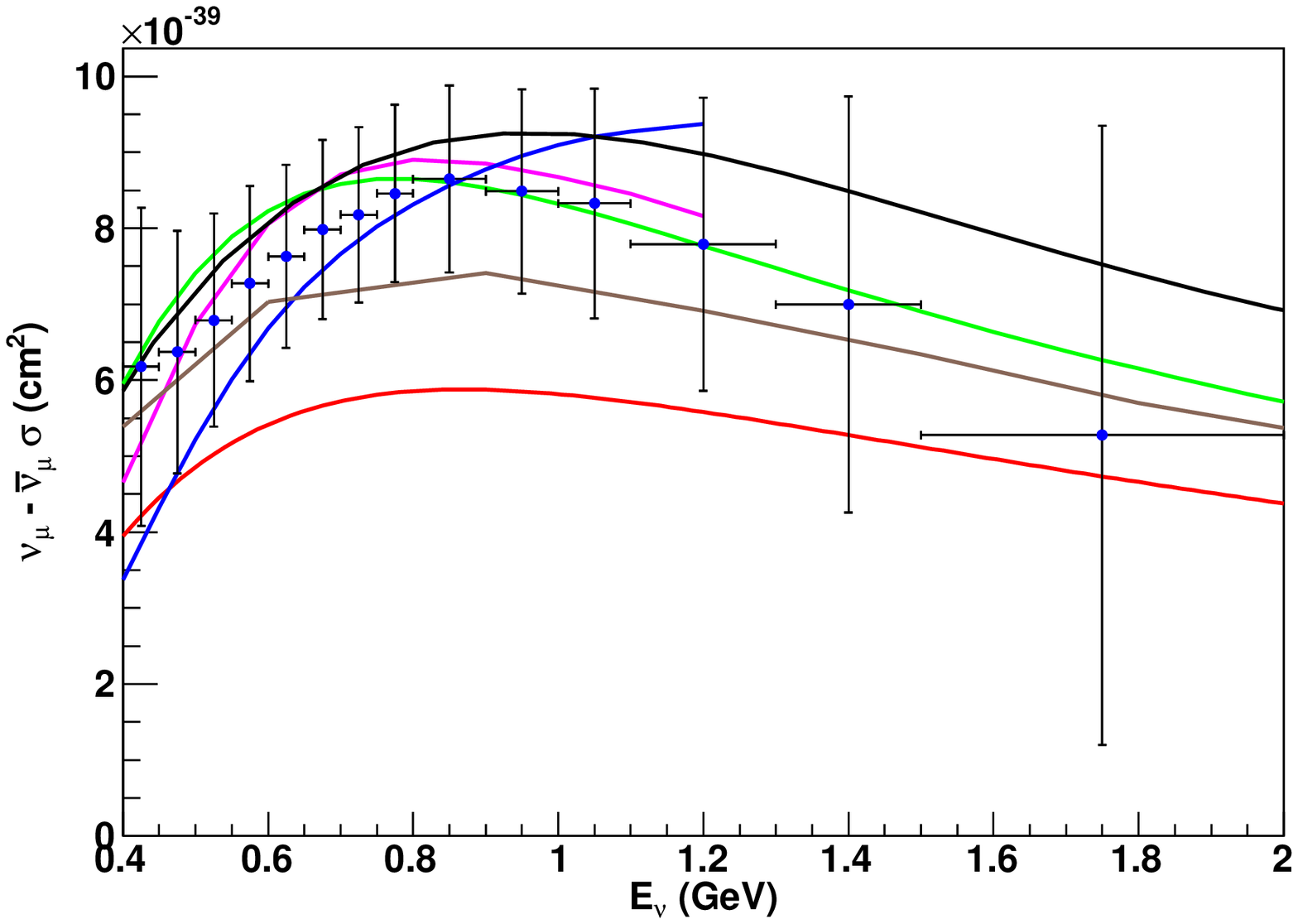} \\
\end{array}$
\vspace*{8pt}
\caption{\label{fig:corrEnu} 
Ratio (left) and difference (right) between MiniBooNE $\numu$ and $\numub$ CCQE 
flux-unfolded total cross section data compared to various predictions. 
Note errors of two data sets are added by quadrature.}
\end{figure}

\begin{figure}[tbp]$
\begin{array}{cc}
\includegraphics[scale=0.33]{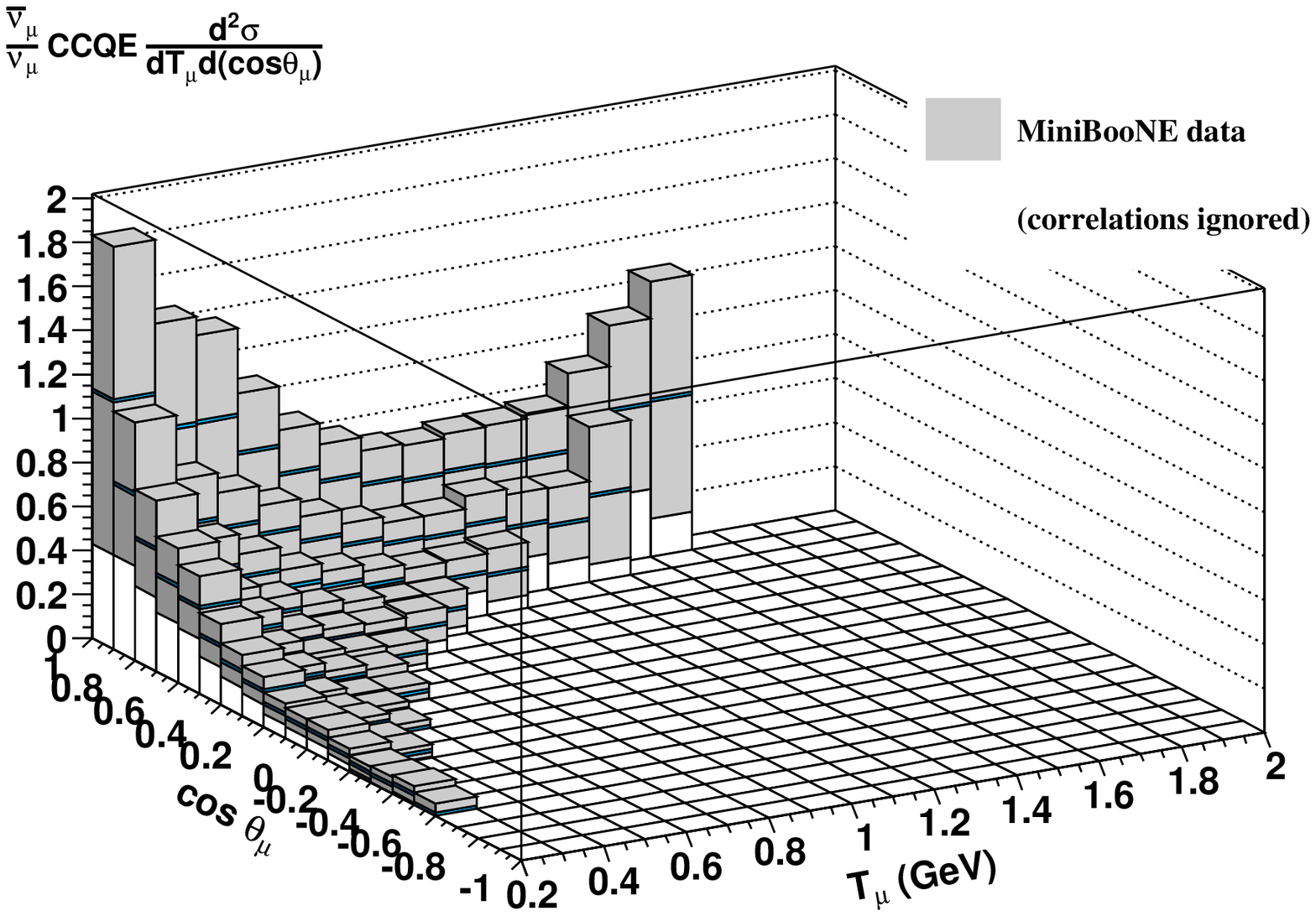} &
\includegraphics[scale=0.33]{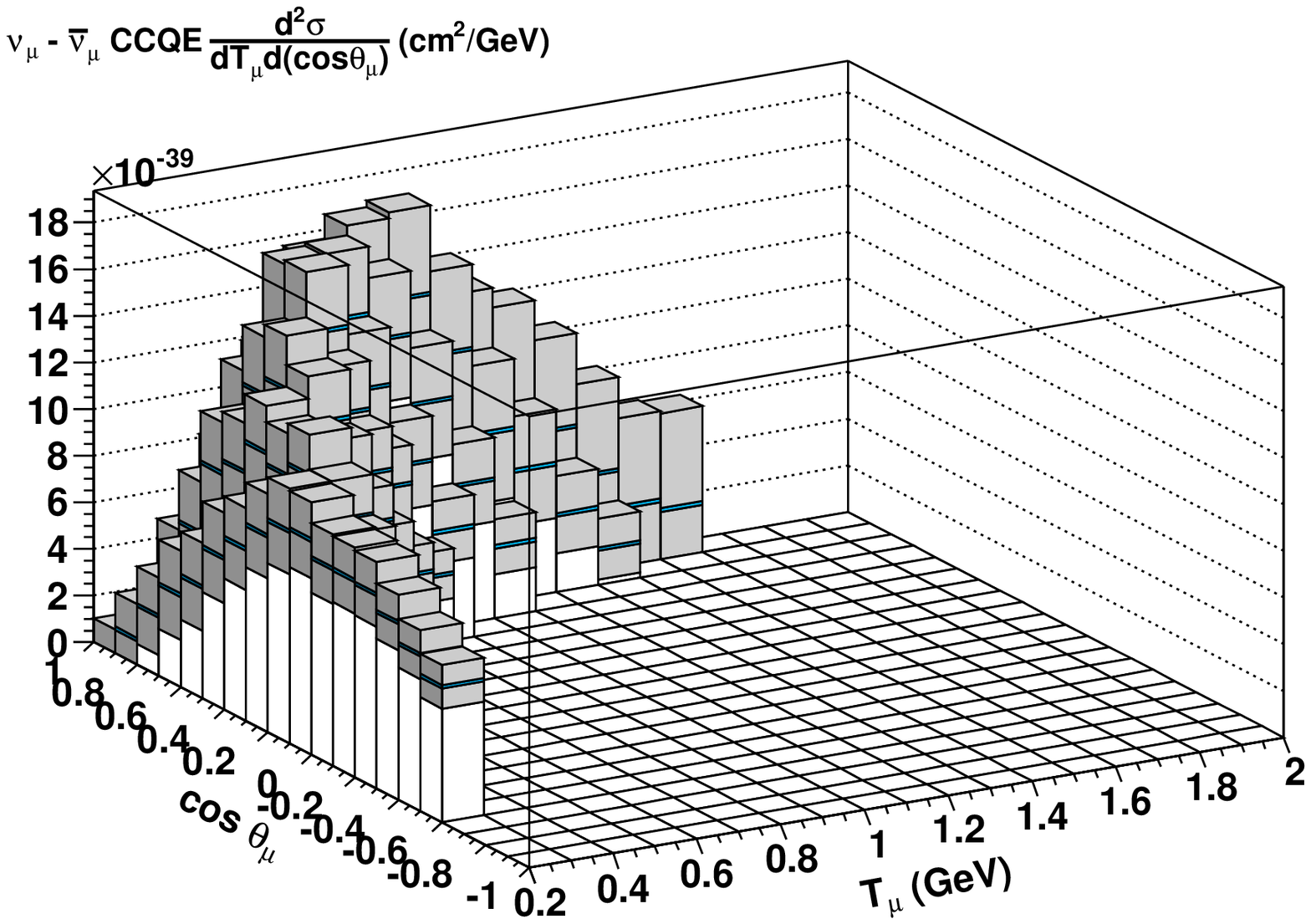} \\
\end{array}$
\vspace*{8pt}
\caption{\label{fig:corrDblDifl} Ratio (left) and difference (right) between MiniBooNE 
$\numu$ and $\numub$ CCQE flux integrated double differential cross section data. 
Note errors of two data sets are added by quadrature.}
\end{figure}

%%Neutrinos probe a different mix of axial and vector cross-section amplitudes compared to antineutrinos, 
%%and any description of the nuclear physics that contribute to these interactions must be verified against data for both processes.  
%%This is of course critical to move towards a rigorous understanding of the underlying mechanisms, 
%%but their accurate description may also soon serve a practical purpose critical 
%%to next-generation neutrino oscillation experiments such as T2K~\cite{T2K_nue2013}, NO$\nu$A~\cite{NOvA_mark}, 
%%and LBNE~\cite{LBNE_snowmass} by providing independent information 
%%for the interaction channel use most prevalently to search for oscillations.  
With the high-statistics MiniBooNE $\numu$ and $\numub$ CCQE measurements described here, 
an opportunity exists to extract even more information by taking the difference or the ratio of them.   
%%In the limit of the nucleon scattering, the difference can be used 
%%to directly measure axial vector form factor $F_A(Q^2)$ interference term,  
%%\beq
%%\frac{d\si^{\nu}}{dQ^2}-\frac{d\si^{\nubar}}{dQ^2}\propto F_A(Q^2)~,\no
%%\eeq
%%This may be true even including nuclear effects. 
%%Because of the isospin nature of the axial term, 
%%the axial vector form factor interference term remains in the cross-section difference.  
%%Furthermore, this term is proportion to the transverse response, 
%%and it is useful to study the transverse nature of the neutrino scattering, too.
%%Because of the isospin nature of the axial vector term, 
%%the difference of cross sections isepolate the term proportion to  
%%the $F_A(Q^2)$ interference term,  
%%\beq
%%\frac{d\si^{\nu}}{dQ^2}-\frac{d\si^{\nubar}}{dQ^2}\prop F_A(Q^2)\cdot(F_1(Q^2)+F_2(Q^2))\cdot R_T~,\no
%%\eeq
%%Here, $F_1(Q^2)$ and $F_2(Q^2)$ are Dirac and Pauli form factors, respectively. 
%%This may be true even including nuclear effects. 
%%Because of the isospin nature of the axial term, 
%%terms proportion to the power one of the axial vector form factor often remain in the difference.  
%%Furthermore, this interference term is proportion to the transverse response, 
%%and it is useful to study some kind of nuclear effects, too.
The cross section ratio is sensitive to the relative contribution of this term in the cross section.
%%out of these data sets by exploiting correlated systematic uncertainties between the two measurements.  
%%Simple difference and ratio analyses between the two results will more stringently test 
%%the various models that predict the size and kinematics of CCQE interactions around 1~GeV. 
%%This analysis is ongoing, but even the quadrature combination of the reported uncertainties offer 
%%some discrimination between various descriptions of these processes. 
For the total cross section, 
these ratio and difference distributions are presented in Figure~\ref{fig:corrEnu}, 
and compared with theoretical predictions from 
Amaro {\it et al.}\cite{Donnelly_anti}, 
Bodek {\it et al.}\cite{Bodek_tem}, 
Martini {\it et al.}\cite{Martini_anti,Martini_ddanti}, 
Nieves {\it et al.}\cite{Nieves_anti}, 
Meucci {\it et al.}\cite{Meucci_anti}, 
along with the RFG and the parameter choice $M_A = 1.35$~GeV. 
While it is interesting to test the overall normalization of the observed $\numu$ and $\numub$ CCQE cross sections 
with more modern ideas of nuclear physics, 
a much more rigorous test is available with comparisons of 
the same quantities with the double-differential cross sections. 
The observed combined measurements of $\numu$ and $\numub$ CCQE are shown in Figure~\ref{fig:corrDblDifl}. 
As this analysis combines information from both $\numu$ and $\numub$ CCQE processes 
while fully exploiting systematic uncertainties, this is the most powerful measurement 
of CCQE interactions possible with the MiniBooNE detector. 

Note that the ratio $\frac{\sigma_{\numu}}{\sigma_{\numub}}$ tests the relative normalization, 
while the difference $\sigma_{\numu} - \sigma_{\numub}$ is sensitive to 
the absolute strength of the cross sections.  
Therefore there is some independent information between these tests, 
and any description  must be consistent with both measurements of the combined cross sections.  
The difference $\numu - \numub$ is 
in principle proportional to the interference term between 
the axial and vector pieces of the formalism for CCQE interactions. 
This is true even including nuclear effects. 
Furthermore, this term is also proportional to the transverse response, 
and it may be useful to study nuclear effects sensitive to the transverse response. 

%We can do even better with these data.  As indicated in Fig.~\ref{fig:corrDblDifl}, the uncertainties on the combined measurements of the double-differential data are produced assuming the systematic uncertainties are completely uncorrelated between the two measurements.  While the backgrounds in the physics samples are mostly unique, possibly strong correlations between detector effects and the HARP hadroproduction data used to predict the $\numu$ and $\numub$ flux may lead to an appreciable level of correlation between the extracted $\numu$ and $\numub$ cross sections.  This would in turn lead to a reduction in uncertainty on the combined measurements shown in Figs.~\ref{fig:corrEnu} and~\ref{fig:corrDblDifl}.  The work is ongoing, 

\section{Conclusion\label{sec:conclusion}}

The CCQE interaction is enormously important in both the history and future of
neutrino physics.  It is a rather simple interaction that has recently
revealed a rich nature. 
MiniBooNE observed structures of this interaction in unprecedented detail, 
and we are waiting further information both from theories and experiments to further
refine this story. 
This is of course critical to move towards a rigorous 
understanding of the underlying mechanisms, 
but their accurate description may also soon serve a practical purpose critical 
to next-generation neutrino oscillation experiments such as T2K~\cite{T2K_nue2013}, 
NO$\nu$A~\cite{NOvA_mark}, and LBNE~\cite{LBNE_snowmass} by providing independent information 
for the interaction channel use most prevalently to search for oscillations.  

%%\section*{Acknowledgement}
%%
%%We thank Rex Tayloe and Sam Zeller for valuable comments.  
%%TK is supported by NSF PHYS-1205175 and STFC-XXXXXXX.

%\section*{References}

\end{document}